\documentclass[%
 pra,
 twocolumn,
 amsmath,amssymb,
 aps,
]{revtex4-2}

\usepackage{bbold}
\usepackage{graphicx}% Include figure files
\usepackage{dcolumn}% Align table columns on decimal point
\usepackage{bm}% bold math
\usepackage{bbm}
\usepackage{hyperref}
\usepackage{braket}
\usepackage[normalem]{ulem}

\newcommand{\dbra}[1]{\langle\!\langle #1\vert}
\newcommand{\dket}[1]{|#1\rangle\!\rangle}

\newcommand{\beginsupplement}{%
	\setcounter{table}{0}
	\renewcommand{\thetable}{S\arabic{table}}%
	\setcounter{figure}{0}
	\renewcommand{\thefigure}{S\arabic{figure}}%
	\renewcommand{\theequation}{S\arabic{equation}}
}

\usepackage{tikz}
\DeclareRobustCommand{\paral}{\tikz{\draw[line width=0.2pt] (0,0) -- (0.15,0.12); 
\draw[line width=0.2pt] (0.03,-0.03) -- (0.18,0.09);}}

\usepackage{xcolor}

\usepackage[normalem]{ulem}

\begin{document}

\preprint{APS/123-QED}

\title{
Cavity-based reservoir engineering for
Floquet-engineered superconducting circuits
}

\author{Francesco Petiziol}
 \email{f.petiziol@tu-berlin.de}
 \affiliation{Technische Universität Berlin, Institut für Theoretische Physik, Hardenbergstraße 36, Berlin 10623, Germany}

\author{Andr{\'e} Eckardt}
 \affiliation{Technische Universität Berlin, Institut für Theoretische Physik, Hardenbergstraße 36, Berlin 10623, Germany}

\date{\today}

\begin{abstract}
Considering the example of superconducting circuits, we show how Floquet engineering can be combined with reservoir engineering for the controlled preparation of target states. Floquet engineering refers to the control of a quantum system by means of time-periodic forcing, typically in the high-frequency regime, so that the system is governed effectively by a time-independent Floquet Hamiltonian with novel interesting properties. Reservoir engineering, on the other hand, can be achieved in superconducting circuits by coupling a system of artificial atoms (or qubits) dispersively to pumped leaky cavities, so that the induced dissipation guides the system into a desired target state. It is not obvious that the two approaches can be combined, since reaching the dispersive regime, in which system and cavities exchange excitations only virtually, can be spoiled by driving-induced resonant transitions. However, working in the extended Floquet space and treating both system-cavity coupling as well as driving-induced excitation processes on the same footing perturbatively, we identify regimes, where reservoir engineering of targeted Floquet states is possible and accurately described by an effective time-independent master equation. We successfully benchmark our approach for the preparation of the ground state in a system of interacting bosons subjected to Floquet engineered magnetic fields in different lattice geometries.
\end{abstract}

\maketitle

\begin{figure}[t]
    \centering
    \includegraphics[width=\linewidth]{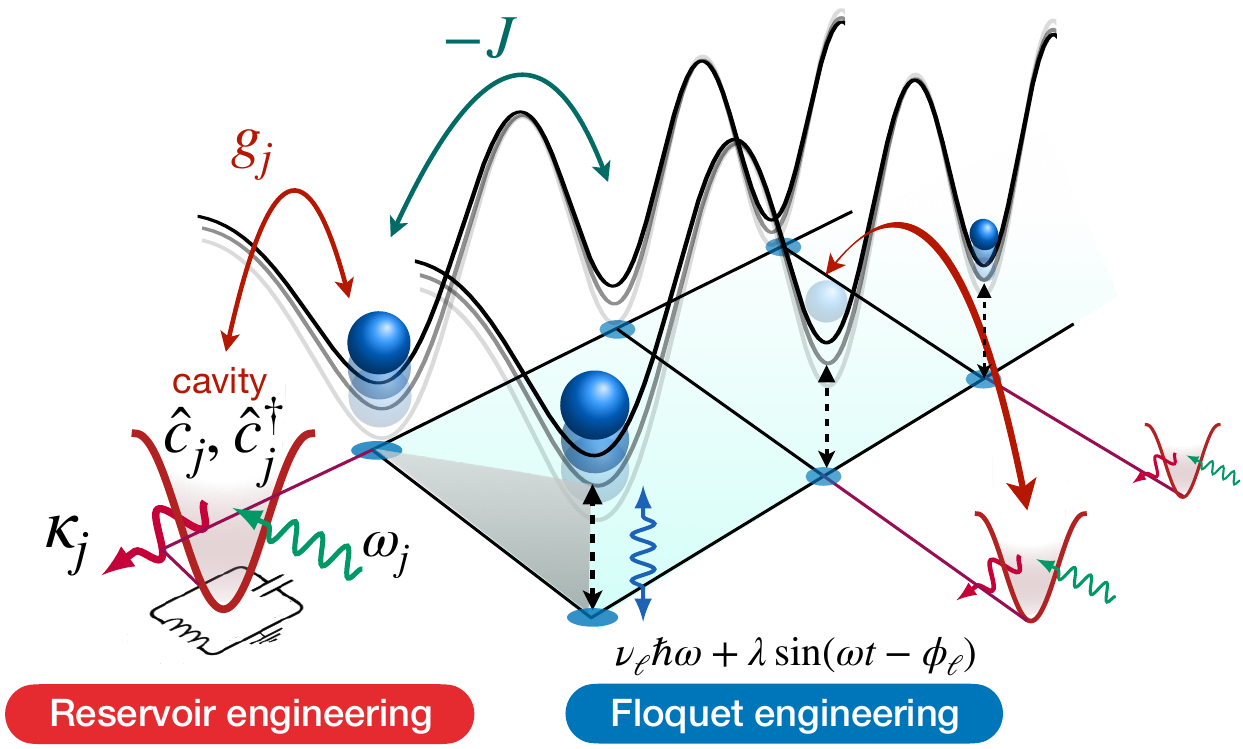}
    \caption{$M$ artificial atoms are coupled to $L$ pumped and leaky cavities. The atoms are periodically driven to Floquet engineer effective Hamiltonians with desired properties, while the cavities induce dissipation in a controlled way.}
    \label{fig1}
\end{figure}

Floquet engineering is a powerful tool for quantum simulation, where time-periodic driving is applied to manipulate the properties of a quantum system~\cite{Goldman2014,Bukov2015, Eckardt2017,Aidelsburger2018,Oka2019,Rudner2020}. It has been applied successfully to engineered quantum systems, such as ultracold atoms in optical lattices~\cite{Lignier2007, Zenesini2009, Struck2011, Struck2012, Atala2014, Jotzu2014, Aidelsburger2015, Tai2017, Wintersperger2020}, photons in optical waveguides~\cite{Rechtsman2013, Maczewsky2017, Ozawa2019}, and superconducting circuits~\cite{Roushan2017b}. Recently, the question has been addressed, whether it is possible to prepare ground (or low-temperature) states of effective Floquet-engineered Hamiltonians by coupling them to a thermal environment~\cite{Breuer1991, Ketzmerick2010, Shirai2015, Shirai2016, Mori2022, Seetharam2015, Iadecola2015}. Here we propose an alternative strategy for dissipative state preparation in Floquet systems based on cavity-assisted reservoir engineering, as it can be realized in superconducting circuits by coupling artificial atoms to pumped leaky resonators. Such an approach is not straightforward, since cavity-based reservoir engineering relies on the so-called dispersive regime, in which the system exchanges excitations with the cavities only virtually. Thus, one has to identify a regime, in which such a virtual change is not spoiled by driving-induced resonant processes. Understanding and avoiding the breakdown of dispersive regimes in circuit QED systems under periodic modulation is a central challenge, as investigated, e.g., in the realization of fast gates and controlled nonlinearities for high-Q cavity modes coupled via driven transmons~\cite{Zhang2019, Zhang2022}. To solve this problem, we describe both atoms and cavities in the extended Floquet space. In this framework, we treat both the system-bath coupling as well as driving-induced excitation processes on equal footing using degenerate perturbation theory. This combined approach contains both the standard perturbative treatment of the dispersive coupling in non-driven systems as well as the high-frequency expansion of isolated periodically-driven systems as limiting cases. But it also includes the interplay of both processes, which has a crucial (potentially detrimental) impact on the open driven dynamics and the preparation of target states. Based on this theory, we formulate driven-dissipative schemes for the preparation of non-trivial states in finite Floquet-engineered flux ladders (exhibiting chiral ground state currents and frustration-induced localization effects). The approach is confirmed via simulations of the full driven-dissipative evolution of atoms and cavities. 

 We consider a two-dimensional array of $M$ artificial atoms (Fig.~\ref{fig1}) in a superconducting circuit~\cite{Roushan2017a, Ma2019, Yan2019, Roushan2017b, Blais2021} described by the Bose-Hubbard Hamiltonian
\begin{align}
    \hat{H}_{\mathrm{S}}(t)  =  \frac{U}{2} \sum_{j=1}^M    \hat{n}_{j}(\hat{n}_{j}-1) 
     -J \sum_{\langle j,j'\rangle} e^{i \theta_{j'j}(t)}\hat{a}_{j'}^\dagger \hat{a}_{j}.
\label{eq:HS}
\end{align}
Here $\hat{a}_j$ and $\hat{n}_j=\hat{a}^\dagger_{j}\hat{a}_j$ denote the bosonic annihilation and number operators for an excitation on site $j$.  The excitations experience an attractive on-site potential $U<0$, corresponding to a level anharmonicity. Moreover, they can tunnel between neighbouring sites with matrix elements of amplitude $J>0$. The time-periodic Peierls phases $\theta_{ij}(t) =  \theta_i(t)-\theta_j(t)$ describe a time-dependent force, which, in a non-rotating reference frame, is described by on-site potentials $v_j= \hbar\dot{\theta}_j=\Delta + \nu_j\hbar\omega+\lambda\sin(\omega t -\varphi_j)$ with integer $\nu_j$. The transition to the rotating frame adopted in Eq.~\eqref{eq:HS} and considered in the following, is accomplished by replacing
$\hat{a}_j\to e^{-i\theta_j(t)} \hat{a}_j$ with $\theta_j(t) = \Delta t/\hbar+ \nu_j \omega t - \lambda \cos(\omega t - \varphi_j)/\hbar\omega$. 

For isolated systems, the motivation for applying such periodic forcing is that in the high-frequency regime, $\hbar\omega\gg |U|, J$ the dynamics is approximately described by an effective time-independent Hamiltonian $\hat{H}_\mathrm{S}^\mathrm{eff}$, with new properties. In leading order, it is obtained as time average $\hat{H}_\mathrm{S}^\mathrm{eff}=\frac{1}{T}\int_0^T \! dt \hat{H}_\mathrm{S}(t)$ over one driving period $T=2\pi/\omega$ (rotating-wave approximation). This gives rise to effective tunneling matrix elements $-J_{jj'}^{\mathrm{eff}}e^{i\theta^\mathrm{eff}_{jj'}}$, with amplitude $J_{jj'}^{\mathrm{eff}}= J\mathcal{J}_{\nu_j-\nu_{j'}}(2\lambda\sin[(\varphi_{j'}-\varphi_j)/2]/\hbar\omega)$,  where $\mathcal{J}_n(\cdot)$ denotes a Bessel function, and
with Peierls phases $\theta^\mathrm{eff}_{jj'} =(\nu_{j'}-\nu_{j})(\varphi_j+\varphi_{j'})/2 $, which can describe artificial magnetic fields~\cite{Eckardt2017}. Such (and similar) Floquet engineering has been employed successfully to experimentally engineer and study interaction-driven phase transitions \cite{Zenesini2009}, kinetic frustration \cite{Struck2011}, topological band structures~\cite{Atala2014, Jotzu2014, Wintersperger2020}, their chiral edge modes~\cite{Rechtsman2013, Maczewsky2017, Roushan2017b}, Aharonov-Bohm cages~\cite{Mukherjee2018}, and two qubit-gates~\cite{Strand2013, Blais2021} in systems of ultracold atoms in optical lattices, optical waveguides, and superconducting circuits.

We will investigate, whether it is possible to employ reservoir engineering for cooling the system into the ground state of $\hat{H}_\mathrm{S}^\mathrm{eff}$. To this end, some of the atoms shall be coupled individually to driven-damped cavities. The open dynamics of the whole system is described by the master equation
 $  d \hat{\rho}/dt = -i [\hat{H}(t), \hat{\rho}]/\hbar + \sum_{\ell=1}^L \kappa_\ell \mathcal{D}[\hat{c}_\ell] \hat{\rho},$ 
where $\mathcal{D}[\hat{c}_\ell]\hat{\rho}=\hat{c}_\ell \hat{\rho} \hat{c}_\ell^\dagger - \frac{1}{2} \hat{c}_\ell^\dagger \hat{c}_\ell \hat{\rho} -\frac{1}{2}\hat{\rho}\hat{c}_\ell^\dagger \hat{c}_\ell$ is a Lindblad dissipator and where the Hamiltonian
$\hat{H}(t)=\hat{H}_\mathrm{S}(t)+\hat{H}_\mathrm{SC}(t)+\hat{H}_\mathrm{C}(t)$ comprises the terms $\hat{H}_\mathrm{C}(t) =    \sum_{j=1}^L \big[\delta_j \hat{c}_j^\dagger \hat{c}_j + \mathcal{E}_j \hat{c}_j^\dagger e^{-i \omega_j t}+\mathcal{E}_j^* \hat{c}_je^{i \omega_j t}]$ and $\hat{H}_{\mathrm{SC}}(t) = \sum_{j=1}^L g_j\big[ e^{-i\theta_{j}(t)}\hat{a}_{j}\hat{c}_j^\dagger + \mathrm{H.c.}]$ that describe the cavities in a frame rotating at frequency $\Delta/\hbar$ and their coupling to the system~\cite{Hacohen2015, Blais2021}. 
The $L\le M$ cavities are described by bosonic annihilation operators $\hat{c}_j$, are detuned by the atoms by $\delta_j$, pumped with strength $\mathcal{E}_j$ at a frequency $\omega_j$, and leak photons at a rate $\kappa_j$. The atoms are enumerated so that the $j$th cavity couples to the $j$th atom with strength $g_j$. The cavity leakage is assumed to be much larger than other decay and dephasing rates in the array, which we thus neglect in the following. The Floquet drive also dresses the array-cavity tunnelling, which hence acquires the phase $\theta_j(t)$.

The atom-cavity Hamiltonian $\hat{H}(t)\equiv  \sum_m \hat{H}_m e^{im\omega t}$
gives rise to Floquet states $|\psi_n(t)\rangle=|u_{n}(t)\rangle e^{-i\varepsilon_{n} t / \hbar } =|u_{n\mu}(t)\rangle e^{-i\varepsilon_{n\mu} t / \hbar }$ with quasienergies $\varepsilon_{n\mu}=\varepsilon_n +\mu\hbar\omega$ that are defined up to integer multiples $\mu$ of $\hbar\omega$ and time-periodic Floquet modes $|u_{n\mu}(t)\rangle =|u_{n}(t)\rangle e^{i\mu\omega t}=|u_{n\mu}(t+T)\rangle\equiv\sum_{\bm{n},\bm{p},m} u_{n\mu}^{(\bm{n}\bm{p}m)} |\bm{n}\bm{p}\rangle e^{im\omega t}$. Here $|\bm{n}\bm{p}\rangle$ denote Fock states with respect to the occupation numbers $\bm{n}=(n_1,\dots,n_M)$ of system excitations and $\bm{p}=(p_1, \dots, p_L)$ of cavity photons and $m$ is an integer Fourier index. In Floquet space~\cite{Sambe1973, Eckardt2017}, spanned by the Floquet-Fock states $|\bm{n}\bm{p}m\rangle\!\rangle$ (representing $|\bm{n}\bm{p}\rangle e^{im\omega t}$ in the original space), the Floquet modes $|u_{n\mu}\rangle\!\rangle=\sum_{\bm{n},\bm{p},m} u_{n\mu}^{(\bm{n}\bm{p}m)}|\bm{n}\bm{p}m\rangle\!\rangle$ are eigenstates with eigenvalue $
\varepsilon_{n\mu}$ of the generalized Hamiltonian $\hat{\mathcal{H}}$ [representing $\hat{H}(t)-i\hbar \partial_t$] with matrix elements
\begin{equation}
\dbra{\bm{n}'\bm{p}'m'}\hat{\mathcal{H}}\dket{\bm{n}\bm{p}m} = \langle\bm{n}'\bm{p}'|(\hat{H}_{m'-m}+\delta_{m'm}m\hbar\omega)|\bm{n}\bm{p}
\rangle.
\end{equation}
The change of $m$ by $\Delta m=m'-m\ne0$, as it is described by the Fourier components $\hat{H}_{\Delta m}$, corresponds to a resonant transition, where the energy changes by $\Delta m$ driving quanta $\hbar\omega$. In turn, processes without such driving-induced energy change are captured by the time-averaged Hamiltonian $\hat{H}_{\Delta m=0}$, which contains $\hat{H}^\text{eff}_\text{S}$. 

\begin{figure}
   \centering
  \includegraphics[width=0.95\linewidth]{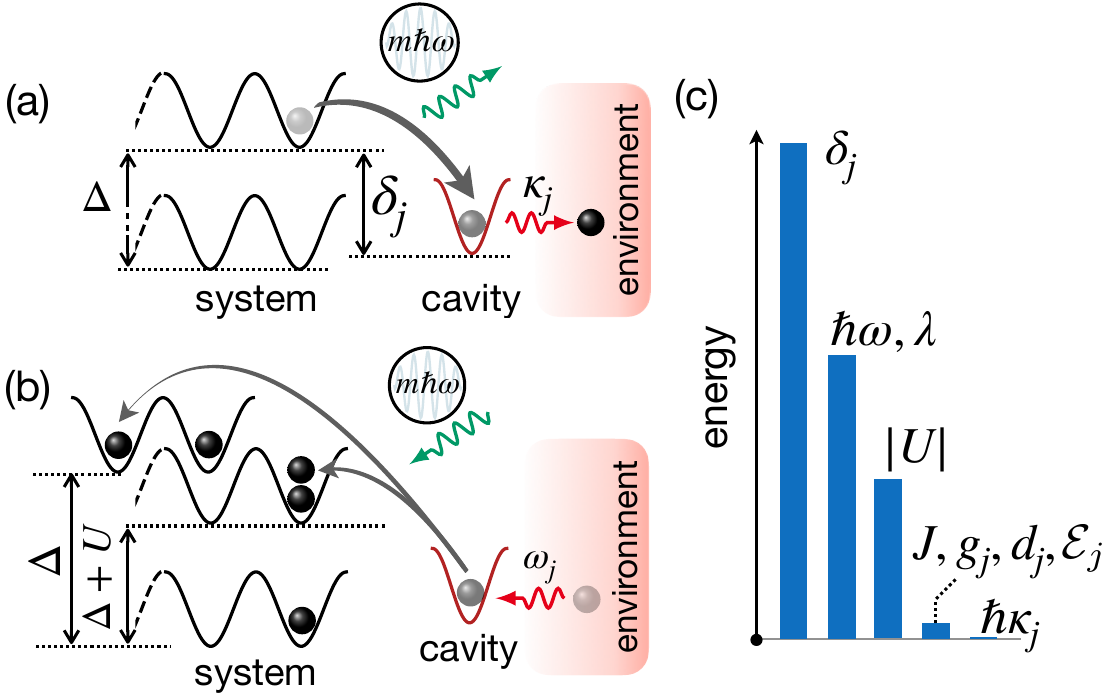}
  \caption{(a)-(b) Examples of unwanted excitation-photon conversion assisted by the Floquet drive. In (a), an excitation tunnels from the array to a cavity releasing an energy $m\hbar\omega$ to the drive, and leaks to the environment being lost. In (b), a photon tunnels from the cavity to the array. {(c)} Fundamental energy scales in the system. }
    \label{fig2}
\end{figure}

For Floquet engineering, we aim at a regime, where $m$ remains a good quantum number, such that the array coherent dynamics is indeed ruled by $\hat{H}_\mathrm{S}^\mathrm{eff}$. For reservoir engineering, we aim at a regime, where the total excitation number $N$ remains a good quantum number, since we strive for quantum simulation at conserved particle number. 
We, thus, aim at a situation, where energy and excitation-number-changing processes can be treated using degenerate perturbation theory, allowing us to block diagonalize $\hat{\mathcal{H}}$ both with respect to $m$ and $N$. This is challenging. While resonances, where $m$ or $N$ change individually are routinely suppressed by making the associated energy costs, the frequency $\hbar\omega$ or the atom-cavity detuning $\delta_j$, respectively, large, it is now also required to avoid resonances where both $m$ and $N$ change. Similar processes are sketched in Fig.~\ref{fig2}(a)-(b), where an excitation can escape [(a)] or be injected [(b)] in the array by exchanging energy $m\hbar\omega$ with the Floquet drive. Although for a weak transverse single-qubit drive as in Ref. \cite{Murch2012} these are very high-order processes, they strongly challenge usual dispersive-regime treatments here and cannot be neglected. Nonetheless, as we will see, the desired regime can be found.

A suitable description is achieved by performing a two-step block-diagonalization of $\hat{\mathcal{H}}$ using van Vleck-type degenerate perturbation theory in Floquet space~\cite{Eckardt2015,SM}. This approach captures corrections arising from processes changing both $m$ and $N$ in second order (except for second-order processes with respect to $m$ alone, which would yield the first correction of $\hat{H}_{\mathrm{S}}^{\mathrm{eff}}$ in a high-frequency expansion and are not relevant here). Such mixed processes would be discarded when time averaging the total Hamiltonian before or after having performed a dispersive-regime transformation, as it is done in Ref.~\cite{Murch2012}. This makes our approach decisive for obtaining correct effective parameters both for the coherent and incoherent dynamics, as confirmed by numerical simulations. We arrive at the effective atom-cavity Hamiltonian~\cite{SM}
\begin{equation}\label{eq:Heff1p}
 \hat{H}_{\mathrm{eff}} = \hat{H}_\mathrm{S}^{\mathrm{eff}} + \hat{H}_\mathrm{C} + \sum_{j=1}^L \big[ \xi_j(\hat{n}_j-1)\hat{n}_j +  \chi_j(\hat{n}_j) \hat{c}_j^\dagger \hat{c}_j\big],
\end{equation}
with functions $\chi_j(\hat{n}_j) =  \widetilde{\chi}_j(\hat{n}_j) - \xi_j(\hat{n}_j)$ and
\begin{align} \label{eq:chixi}
& \widetilde{\chi}_j(\hat{n}_j) = \sum_{m=-\infty}^{+\infty} \frac{g_j ^2 U \mathcal{J}_m^2(\lambda/\hbar\omega)  \hat{n}_j}{[\delta_j + U(\hat{n}_j-1)-m\hbar \omega][\delta_j + U \hat{n}_j - m\hbar\omega]}, \nonumber \\
& \xi_j(\hat{n}_j)  = \sum_{m=-\infty}^{+\infty} \frac{g_j^2 \mathcal{J}_m^2(\lambda/\hbar\omega) }{\delta_j + U \hat{n}_j - m\hbar\omega}.
\end{align}

The coupling to the cavities in Eq.~\eqref{eq:Heff1p} involves only operators preserving the total number of atomic excitations. The prominent role played by the Floquet drive is reflected in the effective tunnelling rate and atom-cavity coupling, which explicitly depend on the driving frequency $\omega$ and amplitude $\lambda$. Equation~\eqref{eq:Heff1p} can only be valid as long as resonances are avoided that make the denominators of Eq.~\eqref{eq:chixi} small. A sketch of the relation between different energy scales in the system, that permit the approximations adopted while being experimentally realistic, is shown in Fig.~\ref{fig2}(c). The starting point is the dispersive regime, where $|\delta_j|\gg g_j$, with a strong nonlinearity $|U|\gg g_j$ that enhances the virtual coupling, see Eq.~\eqref{eq:chixi}. Next $\hbar\omega, \lambda\gg J, g_j$ is chosen to enable Floquet engineering of the tunnelling dynamics, while avoiding resonances. 

Applying the perturbative treatment to the full master equation~\cite{Schnell2021, SM}, an effective dissipator $\mathcal{D}_{\mathrm{eff}}(\hat{\rho})$ is obtained,
$\mathcal{D}_{\mathrm{eff}}(\hat{\rho}) = \sum_{j=1}^L \kappa_j \sum_{m=-\infty}^{+\infty}\mathcal{D}[\hat{c}_{j,m}](\hat{\rho})$,
with
\begin{equation}
    \hat{c}_{j,m}= \hat{c}_j \delta_{m,0} + \frac{g_j \mathcal{J}_m(\lambda/\hbar\omega)}{\delta_j+U \hat{n}_j -m\hbar\omega}\hat{a}_j \ .
\end{equation}
Since the block diagonalization of $\hat{\mathcal{H}}$ mixes atom and cavity degrees of freedom, $\mathcal{D}_{\mathrm{eff}}(\hat{\rho})$ does not involve cavity decay only, but also a small perturbative term $\propto \hat{a}_j$, describing excitation loss from the system. This is analogous to the undriven case in the dispersive regime~\cite{Blais2021}, except for additional driving-induced decay channels with $m\ne0$. %Thus, for the driven system also resonances, where the denominators for finite $m$ vanish, have to be avoided.
The excitation loss is weak in the perturbative regime assumed here, and can be counteracted by postselection~\cite{SM}. In a proof-of-principle implementation, where state tomography in the relevant subspace is accessible, this can be done directly from the estimated density matrix. In large systems, we consider observables that, while being key signatures of the desired effects, carry also information about the total excitation number allowing for postselection. These are site occupations and excitation currents. The former are extracted directly by dispersive readout of the atomic excitations; the latter can be detected as done in cold atom experiments~\cite{Kessler2014, Atala2014}, by biasing the on-site potentials $v_j(t)$ of pairs of neighbouring sites (which is an excitation-conserving process) and measuring the time evolution of site occupations. 

The form of the array-cavity coupling in $\hat{H}_\mathrm{eff}$ puts us in a position, where reservoir engineering for the atoms can be realized. The underlying mechanism is that incoming pump photons that are detuned from the cavity resonance, need to take (or give) energy from (to) the system, before being emitted at the cavity frequency~\cite{Clerk2010, Murch2012, Hacohen2015}. If the detuning matches an energy gap in the atomic system, this induces a corresponding ``dissipative'' transition. This type of quantum bath engineering has been implemented experimentally with superconducting circuits, e.g., for a resonantly driven two-level system and for a three-sites undriven Bose-Hubbard chain~\cite{Murch2012, Hacohen2015}. In our setup, since the relevant atomic Hamiltonian is the Floquet-engineered Hamiltonian $\hat{H}_\mathrm{S}^\mathrm{eff}$, we exploit this mechanism to address transitions among effective eigenstates of the system with artificial magnetic flux.

\begin{figure}[t]
\centering
\includegraphics[width=\linewidth]{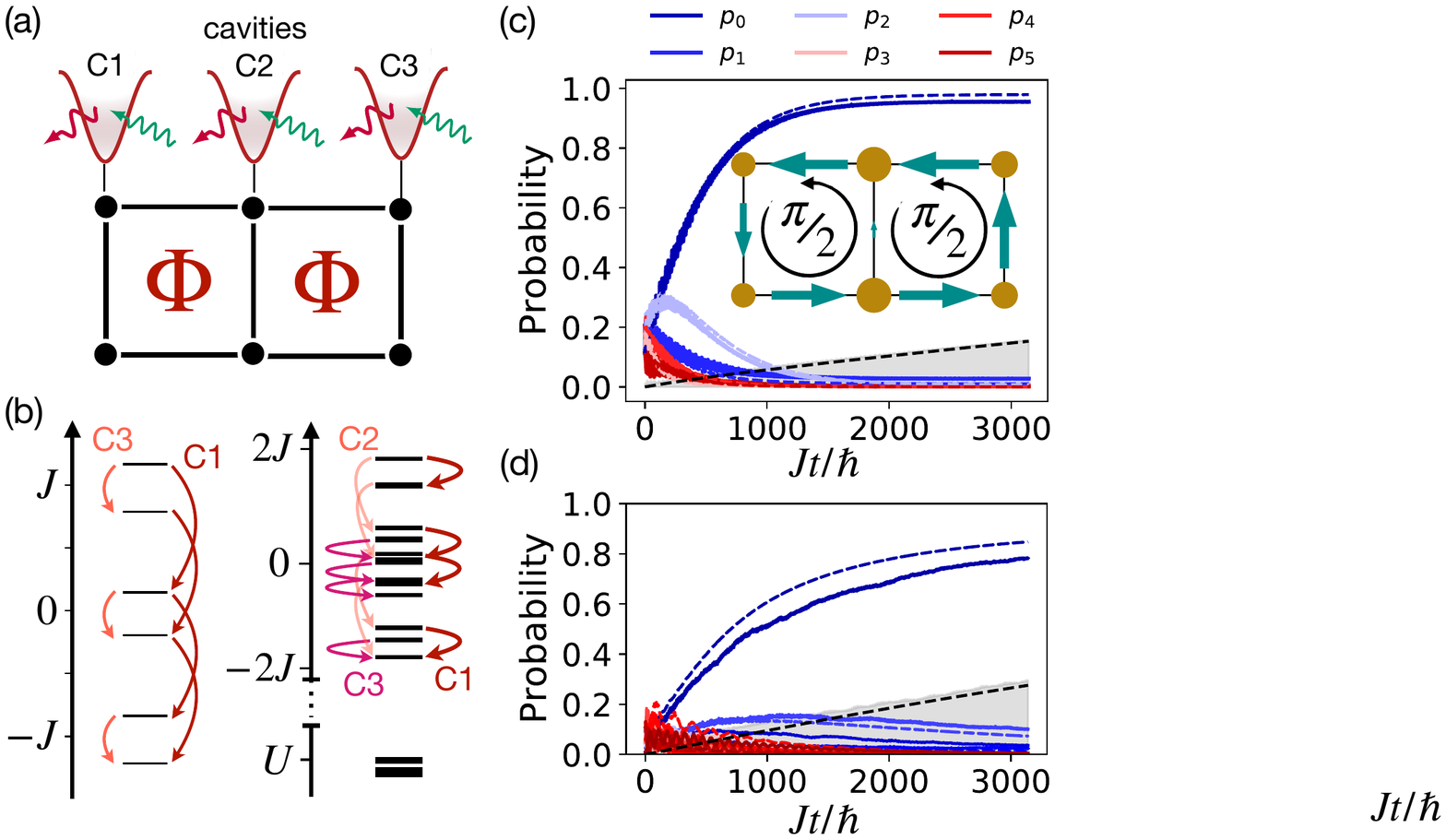}
\caption{(a) Ladder geometry and coupling to the cavities. (b) Single-excitation (left) and two-excitation (right) level structure and transition energies addressed by the cavities (arrows). (c) Stroboscopic evolution of the populations $p_\eta(t)=\bra{\eta} \hat{\rho}(t)\ket{\eta}$ of the eigenstates $\ket{\eta}$ of $\hat{H}_\text{S}^\text{eff}$, for the full driven master equation (solid) and the effective master equation (dashed). The grey shaded area indicates the fraction of discarded states in postselection (black dashed line for the effective model). The parameters are $\Phi=\pi/2$, $\hbar\omega=20J$, $\bm{\delta}/\hbar\omega = (1.76,1.7)$, $U=8J$, $\kappa_1=\kappa_2=0.1J$, $\bm{\mathcal{E}}/J=(1.2, 0.5)$, $g_1=g_2=J$. Inset: The size of the arrows and circles reproduces the current pattern and excitation density in the final state. (d) As (c) but for two excitations in the hard-core bosons subspace. The parameters are $\Phi=\pi/2$, $\hbar\omega=20J$, $\bm{\delta}/\hbar\omega = (1.69,1.68,1.7)$, $U=8J$, $\bm{\kappa}/J=(0.05,0.05,0.05)$, $\bm{\mathcal{E}}/J = (0.6, 1.6, 0.4)$, $\bm{g}/J=(1,1,1)$. }
\label{fig3}
\end{figure}

To design a dissipative path, driving the atoms towards a target eigenstate of $\hat{H}_\mathrm{S}^\mathrm{eff}=\sum_\eta \varepsilon_\eta\ket{\eta}\!\bra{\eta}$, the cavity-pump detunings $d_j$ are set to match different effective energy gaps $\varepsilon_\eta-\varepsilon_{\eta'}$ in the array. Addressing single gaps is possible provided such gaps are larger than the effective atom-cavity coupling $\chi_j^{(\eta\eta')}\sqrt{\bar{n}_{\mathrm{ph},j}}$, where $\bar{n}_{\mathrm{ph},j}$ is the mean photon number in the $j$th cavity and $\chi_j^{(\eta\eta')} = \bra{\eta}\chi_j(\hat{n}_j)\ket{{\eta'}}$ is the matrix element of the atomic coupling operator. The resonant transition rate produced by the $j$th cavity is derived from the effective master equation~\cite{SM} as
\begin{equation}
\Gamma_{\eta \to \eta'}^{(j)} = 4\bar{n}_{\mathrm{ph},j} |\chi_j^{(\eta\eta')}|^2/\hbar^2\kappa_j \ ,
\end{equation}
under the `bad-cavity' condition that the photon leakage is strong compared to the effective coupling, $\hbar \kappa_j \gg \chi_j^{(\eta\eta')} \sqrt{\bar{n}_{\mathrm{ph},j}}$. This condition guarantees a clean monotonic exponential decay from $\ket{\eta}$ to $\ket{\eta'}$, but is not strictly essential: a smaller $\kappa_j$ can still achieve the desired transition, though producing more complex decay dynamics due the atom-cavity system reaching an effective strong-coupling regime~\cite{Murch2012}.

We test the proposed scheme for different systems with Floquet-engineered artificial fluxes. We will compare our effective theory to a full simulation of the time-dependent master equation for both system and cavities, using typical circuit-QED parameters. Given the latter, the determination of the parameters ensuring a successful protocol can be done systematically: while $\hbar\omega$ and the shift $\Delta$ are chosen to be much larger than $J$ and fine-tuned to avoid unwanted resonances, the pump frequencies on the cavities realize the detuning conditions for reservoir engineering. The pump amplitudes are adjusted accordingly to have an average number of one-to-two photons in the cavities, which gave best results in simulations. Further details on parameter choices are given in~\cite{SM}.

We start by considering one excitation in a ladder-type array [Fig.~\ref{fig3}(a)]. The ground state for non-zero magnetic flux $\Phi$ can exhibit chiral currents, flowing unidirectionally along the edges of the ladder~\cite{Atala2014, Hugel2014, Piraud2015, Greschner2015, Wang2021}. Although only one excitation is considered, including subspaces with several excitations and their mutual interactions $U$ is essential, since it influences virtual excitation-number fluctuations exploited for reservoir engineering, see Eqs.~\eqref{eq:Heff1p} and \eqref{eq:chixi}. In the two-plaquettes system of Fig.~\ref{fig3}(a), the ground state is prepared using two cavities which realize a ``cascaded'' cooling configuration along the effective spectrum as depicted in Fig.~\ref{fig3}(b) (left). The driving amplitude $\lambda$ is chosen to give the same effective tunnelling rate $J_{\mathrm{eff}}$ along every lattice bond, while the choice of driving phases and potential offsets yields Peierls phases $r\Phi$ at the $r$th rung~\cite{SM}. The simulated build up of population in the ground state in time is shown in Fig.~\ref{fig3}(c), leading to the final current and density pattern depicted in the inset. The cooling is independent from the initial state. While in Fig.~\ref{fig3}(c) the initial state features the excitation sitting in one atom, an equivalent population build-up is observed also for mixed initial states, such as an infinite-temperature state in the single-excitation manifold. This confirms that the cooling mechanism can reduce the system entropy, in addition to lowering the energy. A cooling scheme in the case of a three-plaquette ladder is reported in \cite{SM}.

The protocol is effective also for multiple excitations. When a second excitation is injected, due to the large on-site interaction $U$, singly- and doubly-excited atoms define two essentially uncoupled subspaces [Fig.~\ref{fig3}(b)-right], corresponding to hard-core bosons and a tightly bound pair, respectively. The former subspace is most interesting (as, for instance, hard-core bosons in finite two-dimensional lattices with homogeneous flux are predicted to give rise to fractional-quantum-Hall(FQH)-type ground states \cite{Soerensen2005, Repellin2020, Wang2022}) and we can use three cavities to cool the system into the hard-core-boson ground state. Starting with two excitations localized in different sites, achievable by exciting two atoms to state $a_{j'}^\dagger a_j^\dagger\ket{0}$ with quick resonant pulses before launching the Floquet drives, the population build-up in time is shown in Fig.~\ref{fig3}(d). 

From Fig.~\ref{fig3}(c) and (d), one can see that the effective master equation reproduces the results of the full model very accurately, that the fraction of discarded data in postselection remains small and does not render the experiment inefficient, and that the target states can be prepared with rather high fidelities. In the Supplemental Material~\cite{SM}, we discuss the geometry of a diamond-chain of corner-sharing rhombic 
plaquettes with flux $\Phi=\pi$ and the preparation of so called Aharonov Bohm cages--- single-excitation states that are localized in a subsystem via destructive interference~\cite{Vidal2000,Mukherjee2018, Kremer2020}.

\begin{figure}
\centering
\includegraphics[width=\linewidth]{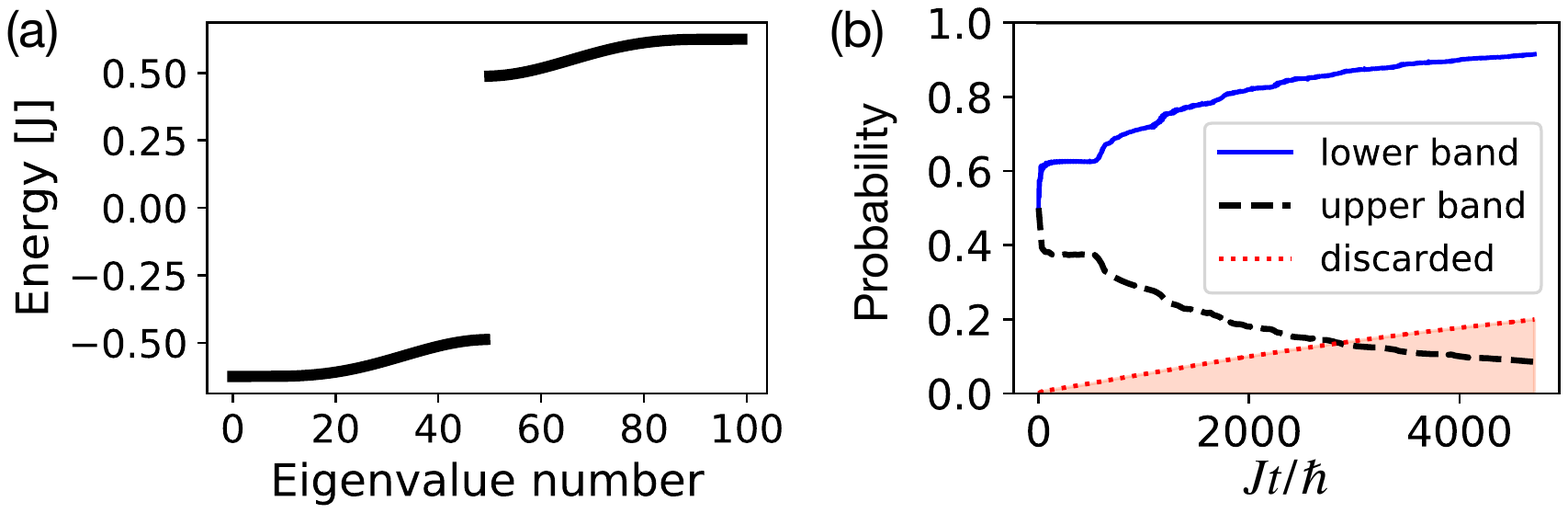}
\caption{(a) Bands in a 100 sites ladder and (b) interband cooling of one excitation using two cavities. Parameters $\Phi=0.8\pi$, $\hbar\omega=20J$, $\lambda \simeq 1.16\hbar\omega$, $d_1=d_2=J$, $\delta_1=\delta_2=34.4J$,  $g_1=g_2=J$,  $\mathcal{E}_1=\mathcal{E}_2=1.4J$. The shaded region indicates the fraction of discarded records in postselection.}
\label{fig4}
\end{figure}

Until now we have addressed the preparation of (zero-temperature) ground states in systems of moderate size. The essential ingredients demonstrated in these examples also provide a perspective for potential applications in larger systems. Namely, the preparation of a gapped ground state, like a Floquet-engineered (topological) band insulator or a correlated fractional Chern insulator, appear possible. Although for a large system, the number of cavities cannot scale as quickly as the number of transitions, multiple transitions can be controlled with a single cavity when they lie in an energy window comparable with the cavity linewidth $\hbar\kappa$~\cite{SM}. This effect can already be appreciated from the two-excitation example of Fig.~\ref{fig3}(b)-(d), where three cavities only are enough to work effectively in the two-excitation subspace (21 states). Another issue with large systems are resonant excitations. Namely the gapped state will most likely be embedded into a continuum of excited states to which it couples resonantly. Although this might limit the capability to resolve the exact ground state, playing a role similar to an effective non-zero temperature, the preparation of low-energy and low-entropy states can still be addressed.

We further exemplify the control of multiple transitions in a larger system, by considering a 100-site square ladder with flux $\Phi=0.8\pi$ and tunnelling along the rungs five times faster than along the legs, whose effective spectrum features two well separated narrow bands [Fig.~\ref{fig4}(a)]. One excitation initialized at a site overlaps with all states in both bands, but can be dissipatively pushed into the lower band using two cavities only, coupled to the two ends of the upper leg. We demonstrate this successfully by performing simulations with the effective master equation [Fig.~\ref{fig4}(b)]. This example further highlights that Floquet-dissipative schemes can also be used to project a system into a (quasi)energetically well separated subspace, opening potential applications for autonomous quantum error correction ~\cite{Terhal2015, Gertler2021} and the preparation of gapped ground states (such as FQH states \cite{Soerensen2005, Repellin2020, Liu2021, Wang2022}).
In view of the latter, particularly exciting is the combination of artificial gauge fields~\cite{Goldman2014,Aidelsburger2018} or geometric frustration~\cite{Eckardt2010, Struck2011}, as it can be achieved using Floquet engineering, with an interaction-induced hard-core constraint. Such hard-core interactions can not only be used to mimic fermionic behavior in 1D, but are also predicted to stabilize fractional-Chern-insulator states in topologically non-trivial lattice systems as well as spin-liquid-like states. Finally, the theory developed here can find straightforward application also in different platforms where Floquet engineering has been employed, such as quantum gas microscopes, whenever coupling to driven cavity-like modes is possible. 
 
 This research was funded by the German Research Foundation (DFG) via the Collaborative Research Center (SFB) 910, under project number 163436311, and the Research Unit FOR 2414, under Project No. 277974659.

 \let\oldaddcontentsline\addcontentsline% Store \addcontentsline
\renewcommand{\addcontentsline}[3]{}% Make \addcontentsline a no-op
\bibliography{biblio}
\let\addcontentsline\oldaddcontentsline% Restore \addcontentsline

\cleardoublepage

\beginsupplement
\setcounter{page}{1}
\onecolumngrid
\begin{center}
{\bf \large Supplemental Material to} \\
\vspace{0.2cm}
{\bf \large Cavity-based reservoir engineering for Floquet-engineered superconducting circuits} \\
\vspace{0.4cm}

{ Francesco Petiziol and Andr\'e Eckardt}\\

%\vspace{0.3cm}

{\itshape
Technische Universität Berlin, Institut für Theoretische Physik, Hardenbergstraße 36, Berlin 10623, Germany}
\end{center}

\tableofcontents

\section{Effective master equation.}

\noindent In this section, the effective Hamiltonian of Eq. (3) in the main text and the effective master equation are derived.

\subsection{Rotating frames.} 

We start by deriving the form of the Hamiltonian given in Eq. (1) in the main text. The Hamiltonian of the atom array reads 
$\hat{H}_\mathrm{S}(t) = \hat{H}_\mathrm{op}(t) + \hat{H}_{\mathrm{int}} + \hat{H}_{\mathrm{tun}}$ with
\begin{align} \label{eq:start}
    \hat{H}_{\mathrm{op}}(t) = &  \sum_{\ell=1}^M  \hbar\dot{\theta}_\ell(t) \hat{n}_{\ell}, \quad \hat{H}_{\mathrm{int}} = \frac{U}{2} \sum_{\ell=1}^M \hat{n}_{\ell}(\hat{n}_{\ell}-1), \quad \hat{H}_{\mathrm{tun}} = -J \sum_{\langle \ell,\ell'\rangle}  (\hat{a}_{\ell'}^\dagger \hat{a}_{\ell} + \hat{a}_{\ell'} \hat{a}_{\ell}^\dagger),
\end{align}
where $\langle \ell, \ell'\rangle$ indicates ordered pairs of nearest neighbours in the lattice. The periodically modulated site potential has sinusoidal shape, $\hbar\dot{\theta}_\ell(t) =\nu_\ell \hbar \omega + \lambda \sin(\omega t - \varphi_\ell).$ 
The offsets $\nu_\ell\hbar\omega$, where $\nu_\ell\in\{0,1\}$ in our applications, are needed for Floquet engineering artificial magnetic fields. The Hamiltonian $H_{\mathrm{S}}(t)$ is written in a frame rotating at frequency $\Delta/\hbar$, where $\Delta$ is a static atomic potential equal for all atoms.
The Hamiltonian for the $L$ cavities in the frame rotating at the same frequency $\Delta/\hbar$ reads
\begin{equation}
    \hat{H}_\mathrm{C}(t)  = \sum_{\ell=1}^L \Big(\delta_\ell \hat{c}_\ell^\dagger \hat{c}_\ell + \mathcal{E}_\ell \hat{c}_\ell^\dagger e^{-i\omega_\ell t} + \mathcal{E}^*_\ell  \hat{c}_\ell e^{i \omega_{\ell} t }\Big),
\end{equation}
such that $\delta_\ell$ represents the detuning between the cavity frequencies and the static atomic potential $\Delta$.
The quantity $\omega_\ell$ is then the detuning between the frequency of the cavity pump on the $\ell$th cavity from $\Delta$.
 The atom-cavity interaction is given by the Hamiltonian $ \hat{H}_{\mathrm{SC}} = \sum_{\ell=1}^L g_\ell (\hat{a}_{\ell}\hat{c}_\ell^\dagger + \hat{a}_\ell^\dagger \hat{c}_\ell)$, characterized by the coupling strengths $g_\ell$. The dynamics of the combined cavities-atoms system is then described by the Lindblad master equation
\begin{equation}\label{eq:mastereq}
    \frac{d \hat{\rho}}{dt} = -\frac{i}{\hbar} [\hat{H}_\mathrm{S}(t) + \hat{H}_\mathrm{C}(t) + \hat{H}_\mathrm{SC}, \hat{\rho}] + \sum_{\ell=1}^L \kappa_\ell \mathcal{D}[\hat{c}_\ell] \hat{\rho} \ ,
\end{equation}
where $\mathcal{D}[\hat{c}_\ell]\hat{\rho}$ is a Lindblad dissipator describing cavity photon leakage, 
 $   \mathcal{D}[\hat{c}_\ell]\hat{\rho} = \hat{c}_\ell \hat{\rho} \hat{c}_\ell^\dagger - \frac{1}{2} \hat{c}_\ell^\dagger \hat{c}_\ell \hat{\rho} -\frac{1}{2}\hat{\rho}\hat{c}_\ell^\dagger \hat{c}_\ell.$
As discussed in Sec. \ref{sec:numerical_simulations}, this is the master equation used to simulate the full driven dynamics of system and cavities. 
To arrive at Eq. (1), we represent the master equation in the time-dependent frame comoving with the Floquet drive and the potential offsets $\nu_\ell \hbar \omega$, {\it i.e.}, such that these terms are canceled from the comoving-frame Hamiltonian. The comoving frame is defined by the unitary transformation
\begin{subequations}\label{eq:R}
\begin{align}
    R(t) & = \exp\Big\{-i\sum_{\ell=1}^M \int_0^t  \dot{\theta}_\ell(t) \hat{n}_{\ell}\Big\}, \nonumber\\
    & = \exp\Big\{- i \sum_{\ell=1}^M \Big( \nu_\ell \omega t + \frac{\lambda}{\hbar\omega}\big[\cos(\varphi_\ell) - \cos(\omega t - \varphi_\ell)\big]\Big)\hat{n}_{\ell} \Big\}.
\end{align}
\end{subequations}
Note that $R(0)=\mathbb{1}$ and that $R(t+T)=R(t)$ is time-periodic. Hence, $R(t)$ maps to the identity at stroboscopic times $nT$, $n\in \mathbb{N}$, and thus the rotating frame coincides with the laboratory frame at such times. The comoving-frame Hamiltonian $R^\dagger(t) \hat{H}(t)R(t)-i\hbar R^\dagger(t) \partial_t R(t)$ reads
\begin{equation}\label{eq:wideH}
    \hat{H}(t) = \hat{H}_{\mathrm{int}} + \hat{H}_{\mathrm{C}}(t) + \hat{H}_{\mathrm{tun}}(t)  + \hat{H}_{\mathrm{SC}}(t),
\end{equation}
with
\begin{equation}\label{eq:Hwidepieces}
\qquad \hat{H}_{\mathrm{tun}}(t) = -J \sum_{\langle\ell, \ell'\rangle}  e^{i \theta_{\ell'\ell}(t)} \hat{a}^\dagger_{\ell'}\hat{a}_\ell + \mathrm{h.c.}, \qquad \hat{H}_{\mathrm{SC}}(t) = \sum_{\ell=1}^L g_\ell e^{-i \theta_{\ell}(t)} \hat{a}_\ell \hat{c}^\dagger_\ell + \mathrm{h.c.},
\end{equation}
The time-dependent Peierls phases $\theta_\ell(t)$ and $\theta_{\ell\ell'}(t)=\theta_\ell(t) - \theta_{\ell'}(t)$ are given by
\begin{subequations}\label{eq:peierls}
\begin{align}
& \theta_\ell(t) = \nu_\ell \omega t-\frac{\lambda}{\hbar\omega} \cos(\omega t - \varphi_\ell),\\
& \theta_{\ell\ell'}(t) =  \frac{2\lambda}{\hbar\omega} \sin\left(\frac{\varphi_{\ell}-\varphi_{\ell'}}{2} \right)\sin\left(\omega t - \frac{\varphi_{\ell}+\varphi_{\ell'}}{2} \right) + \nu_{\ell\ell'} \omega t.
\end{align}
\end{subequations}
with $\nu_{\ell\ell'} = \nu_{\ell} - \nu_{\ell'}$. Note that, slightly differently from the main text, the frequency $\Delta/\hbar$ is not included into $\theta_\ell(t)$ here, since Eq.~\eqref{eq:start} was already written in a frame rotating at $\Delta/\hbar$. The Hamiltonian $\hat{H}(t)$ of Eq.~\eqref{eq:wideH} thus corresponds to $\hat{H}_\mathrm{S}(t) + \hat{H}_\mathrm{C}(t)+\hat{H}_\mathrm{SC}(t)$ in the main text.  
In the transformation, we have absorbed time-independent phase factors $e^{-i \lambda\cos(\varphi_\ell)/\hbar\omega}$ into the operators,
\begin{equation}
e^{i \lambda\cos(\varphi_\ell)/\hbar\omega}\hat{a}_\ell^\dagger \to \hat{a}_\ell^\dagger,
\end{equation}
 making a gauge transformation. In the following, it will be assumed that the cavities are coupled to sites for which $\nu_\ell=0$, so we will set $\nu_\ell=0$ in $\hat{H}_{\mathrm{SC}}(t)$. The dissipator remains unmodified under the change of rotating frames.

\begin{figure}
\includegraphics[width=0.7\linewidth]{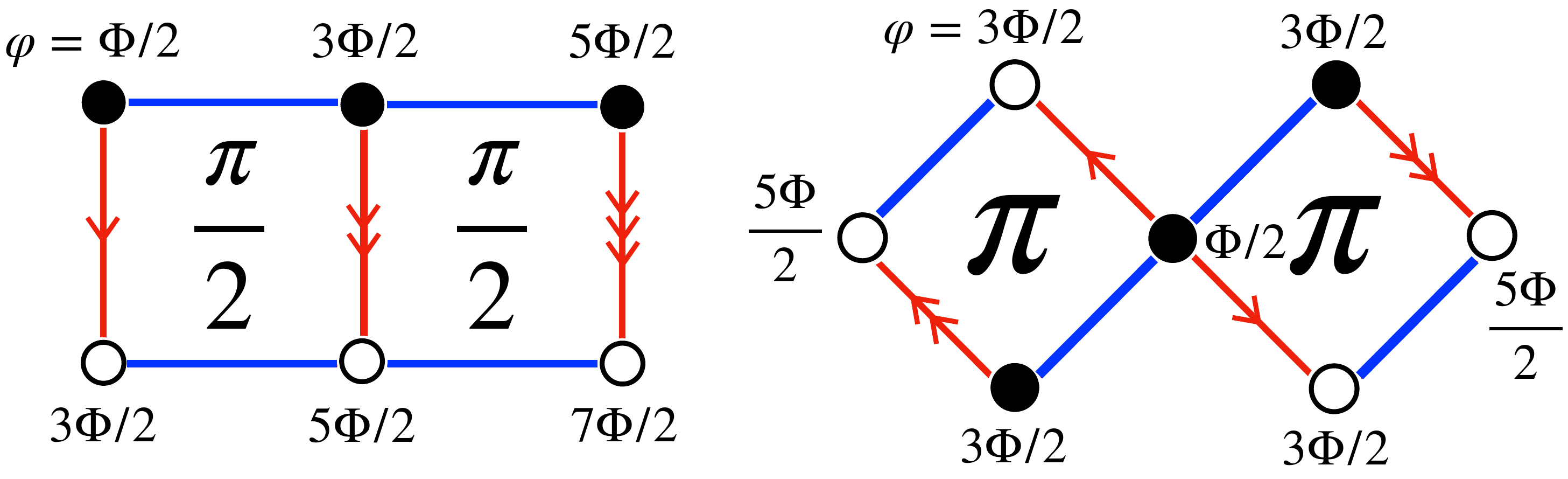}
\caption{Driving phases and resulting coupling patterns. The phases $\varphi_\ell$ of the Floquet drives are indicated at the corresponding lattice sites. Blue (red) links indicate a tunneling rate renormalized with the Bessel function $\mathcal{J}_0$ ($\mathcal{J}_1$). Empty circles indicate sites where there is a potential offset of $\hbar\omega$. Each arrow indicates a phase factor $e^{i\Phi}$ representing the Peierls phases engineered with the Floquet drives.}
\label{fig:drive_phases}
\end{figure}

\subsection{ Artificial magnetic flux.} 

The Floquet engineering protocols considered in this work aim at realizing effective magnetic fields, which enable the study of time-reversal symmetry breaking and frustration with system excitations, which would not respond to a physical magnetic field. We briefly review here, how the periodic drives yield an artificial magnetic flux in our setup ~\cite{Eckardt2017, Aidelsburger2018}. Neglecting the cavities for the moment, high-frequency Floquet engineering of the tunneling dynamics is possible if the Floquet drives act on a timescale which is much faster than the tunneling rate $J$, namely $\hbar \omega \gg J$ and $\lambda\sim\hbar\omega$. The artificial flux is introduced by shaping with the drives the time averaged tunneling Hamiltonian, $\frac{1}{T}\int_0^T \!dt\ \hat{H}_{\mathrm{tun}}(t).$ From Eq.~\eqref{eq:Hwidepieces} and using the generating function relation for the Bessel functions of the first kind,
$e^{\frac{z}{2}(t - t^{-1})} = \sum_{m=-\infty}^\infty \mathcal{J}_m(z) t^m$, for $ z\in \mathbb{C}, t \in \mathbb{C}\backslash\{0\}$, this time average gives
\begin{equation}
\frac{1}{T}\int_0^T\!dt\ \hat{H}_{\mathrm{tun}}(t)  = - \sum_{\langle \ell, \ell'\rangle} J^{\mathrm{eff}}_{\ell\ell'} e^{i\theta^{\mathrm{eff}}_{\ell\ell'}} \hat{a}_{\ell'}^\dagger \hat{a}_{\ell} + \mathrm{h.c.}\ ,
\end{equation}
with effective tunneling rate as given in the main text,
\begin{equation}
J^{\mathrm{eff}}_{\ell\ell'} = J  \mathcal{J}_{\nu_{\ell\ell'}}\left( \frac{2\lambda}{\hbar\omega} \sin\big[(\varphi_{\ell'}-\varphi_{\ell})/2\big] \right),
\end{equation}
and effective phases $\theta^{\mathrm{eff}}_{\ell\ell'} = \nu_{\ell'\ell} \frac{\varphi_{\ell'}+\varphi_{\ell}}{2}.$ 
Beyond renormalizing the tunneling rate, the Floquet drives then permit one to shape the complex tunneling phases $\theta^{\mathrm{eff}}_{\ell\ell'}$. 
In particular, note that when two neighbouring sites $\ell, \ell'$ have equal on-site potential offset, $\nu_{\ell\ell'} = 0$, then the effective phase is zero, while it is non-zero whenever $\nu_{\ell\ell'} \ne 0$. In the examples used in the numerical simulations and presented in the main text, the driving phases and amplitude are chosen to give the same renormalized tunneling strength, at chosen artificial flux $\Phi$. This is 
\begin{equation}
J^{\mathrm{eff}}_{\ell\ell'}/J = \mathcal{J}_0\left( \frac{2\lambda}{\hbar\omega}\sin(\Phi/2)\right) = \mathcal{J}_1\left( \frac{2\lambda}{\hbar\omega}\sin(\Phi/2)\right),
\end{equation}
with $\lambda$ given, by the condition that $\mathcal{J}_0$ and $\mathcal{J}_1$ have the same value, $\lambda/\hbar\omega \simeq 0.72/\sin(\Phi/2)$. For flux $\Phi=\pi/2$, this yields the effective rate $J_{\ell\ell'}^{\mathrm{eff}}\simeq0.55J$. The pattern of driving phases and the resulting pattern of couplings and Peierls phases for the two-plaquette ladder and rhombic lattices discussed in the article is shown in Fig.~\ref{fig:drive_phases}.

\subsection{``Floquet-dispersive'' regime.}
In the first perturbative step discussed in the text, the goal is to block-diagonalize the system-cavity coupling in the Floquet Hamiltonian with respect to the number $N$ of atomic excitations, thus extending the usual dispersive regime of circuit QED~\cite{Blais2021} to the driven problem. This allows us to identify regimes where system and cavities exchange energy without exchanging excitations. To this end, the Hamiltonian of Eq.~\eqref{eq:wideH} is split into an unperturbed part $\hat{H}_\mathrm{unp}$, which commutes with $\hat{N}=\sum_j \hat{n}_j$, and a perturbation $\hat{V}$ which we will treat using van Vleck perturbation theory \cite{Shavitt1980}. The unperturbed part is given by the time independent terms,
\begin{equation} \label{eq:Hunp}
 \hat{H}_{\mathrm{unp}} =  \frac{U}{2}\sum_{j=1}^M \hat{n}_j(\hat{n}_j-1) + \sum_{j=1}^L \delta_j \hat{c}^\dagger_j \hat{c}_j \ , 
 \end{equation}
where for the moment we do not include the cavity pump terms in $\hat{H}_\mathrm{C}$ (see Section ``Cavity pumps'' below).
The perturbation reads
\begin{equation}\label{eq:perturb}
    \hat{V}(t) = \hat{H}_{\mathrm{tun}}(t) + \hat{H}_{\mathrm{SC}}(t) \ .
\end{equation}
As discussed in the main text, the perturbation contains unwanted Floquet $m$-photon processes that can lead to excitation loss or creation in the system. The Floquet Hamiltonian $\hat{\mathcal{H}}$ in the extended space \cite{Sambe1973, Eckardt2015} then reads
\begin{equation}
    \hat{\mathcal{H}} = \hat{\mathcal{H}}_{\mathrm{unp}} + \hat{\mathcal{V}} \ ,
\end{equation}
with $\hat{\mathcal{H}}_{\mathrm{unp}}$ and $\hat{\mathcal{V}}$ defined as follows. The unperturbed part $\hat{\mathcal{H}}_{\mathrm{unp}}$, representing $\hat{H}_\mathrm{unp}$ of Eq.~\eqref{eq:Hunp} in the extended space, is defined as
\begin{equation}\label{eq:Qunp}
    \hat{\mathcal{H}}_{\mathrm{unp}} =\sum_{\bm{n p} } \sum_{m,m'=-\infty}^{+\infty} \dket{\bm{n}\bm{p} m} \varepsilon_{\bm{n} \bm{p} m} \dbra{\bm{n}\bm{p} m} \ ,
\end{equation}
where the unperturbed eigenstates $\dket{\bm{n}\bm{p}m} $ are labeled by: (i) the number $m$ of drive quanta $\hbar\omega$ in the Floquet drive (ii) $\bm{n} = \{n_s\}_{s=1,\ldots,M}$, the set of excitation numbers in the array (iii) $\bm{p}=\{p_s\}_{s=1,\ldots,L}$, the number of photons in each cavity mode. In this section, we will use a calligraphic notation $\hat{\mathcal{O}}$ to indicate the extended-space representation of an operator $\hat{O}$.
The unperturbed quasienergies explicitly read
\begin{align}\label{eq:upenergies}
  \varepsilon_{\bm{n}\bm{p} m} = & \bra{\bm{n} \bm{p}}\hat{H}_{\mathrm{unp}} \ket{\bm{n}\bm{p}} + m\hbar\omega\ , \nonumber\\
  = &  \frac{U}{2}\sum_{\ell=1}^M n_\ell(n_\ell-1)+ \sum_{\ell=1}^L \delta_\ell p_\ell + m\hbar \omega \ .
\end{align}
The perturbation $\hat{\mathcal{V}}$ reads 
\begin{equation}\label{eq:pert_x}
\hat{\mathcal{V}} = \sum_{m,m'} \ket{m}_{\! F} \Big(\hat{H}_{\mathrm{tun},m-m'} +\hat{H}_{\mathrm{SC},m-m'}  \Big) {}_F\!\bra{m'},
\end{equation}
where we introduced the convenient short-hands
\begin{eqnarray}
    \ket{m}_{\! F} \equiv \sum_{\bm{p}\bm{n}} |\bm{p}\bm{n}m\rangle\!\rangle \langle\bm{p}\bm{n}|,
    \qquad
    {}_{F}\!\bra{m'} \equiv \sum_{\bm{p}'\bm{n}'} |\bm{p}'\bm{n}'\rangle\langle\!\langle \bm{p}'\bm{n}'m'| .
\end{eqnarray}
From Eqs.~\eqref{eq:Hwidepieces} and \eqref{eq:peierls}, the Fourier components appearing in Eq.~\eqref{eq:pert_x} are
\begin{subequations}
\begin{align}
    \hat{H}_{\mathrm{tun}, m}= & - \sum_{\langle \ell,\ell' \rangle}  \big(J_{\ell\ell',m} \hat{a}_{\ell'}^\dagger \hat{a}_{\ell} + J_{\ell'\ell,m} \hat{a}_{\ell'} \hat{a}_{\ell}^\dagger \big)\ , \\ 
    \hat{H}_{\mathrm{SC},m} = & \sum_{j=1}^L g_{j,m} [\hat{a}_j \hat{c}_j^\dagger +(-1)^m \hat{a}_j^\dagger \hat{c}_j] \ , \label{eq:HAR_m}
\end{align}
\end{subequations}
where we have defined
\begin{subequations}\label{eq:Jjmgjm}
\begin{align}
J_{\ell\ell',m} = & J\mathcal{J}_{m-\nu_{\ell'\ell}}\left(\frac{2\lambda}{\hbar\omega}\sin\big[(\varphi_{\ell'} - \varphi_{\ell})/2\big] \right) e^{-i(m-\nu_{\ell'\ell})(\varphi_{\ell'} + \varphi_{\ell})/2},\\
g_{j,m} =& g_j \mathcal{J}_m\Big( \frac{\lambda}{\hbar\omega}\Big)e^{i m(\pi/2 - \varphi_{j})}.
\end{align}
\end{subequations}
We now decompose $\hat{\mathcal{V}} = \hat{\mathcal{V}}_{\mathrm{D}}+\hat{\mathcal{V}}_{\mathrm{OD}}$ into its block-diagonal ($\hat{\mathcal{V}}_{\mathrm{D}}$) and block-off-diagonal ($\hat{\mathcal{V}}_{\mathrm{OD}}$) part with respect to the total excitation number $N$ in the array. The two components are then
\begin{subequations}\label{eq:Vparts}
\begin{align}
    & \hat{\mathcal{V}}_{\mathrm{D}} = \sum_{m,m'=-\infty}^{+\infty}\ket{m}_{\! F} \Big(\hat{H}_{\mathrm{tun},m-m'}   \Big){}_F\!\bra{m'} \ ,\label{eq:olVD} \\
   &  \hat{\mathcal{V}}_{\mathrm{OD}} = \sum_{m,m'=-\infty}^{+\infty}\ket{m}_{\! F} \Big(\hat{H}_{\mathrm{SC},m-m'}  \Big) {}_F\!\bra{m'} \ .\label{eq:Vpartsb}
\end{align}
\end{subequations}
A transformation $
    \hat{\mathcal{U}} = \exp(-\hat{\mathcal{G}}) = \exp\left(-\sum_{k=1}^{+\infty}  \hat{\mathcal{G}}^{(k)}\right)$
is searched which block-diagonalizes the array-cavity coupling in $\hat{\mathcal{H}}$ with respect to $N$,
\begin{equation}\label{eq:bdiag1}
    \hat{\mathcal{U}}^\dagger \hat{\mathcal{H}}\hat{\mathcal{U}} = \hat{\mathcal{W}}\ ,
\end{equation}
with $\hat{\mathcal{W}}$ block-diagonal.
Assuming that a perturbative expansion is justified, namely if the unperturbed gaps are much larger than the coupling to the perturbation, each generator $\hat{\mathcal{G}}^{(k)}$ will be of order $k$ in the perturbative parameter and one can impose the block-diagonalization up to a desired order~\cite{Eckardt2015}. The relevant coupling in $\hat{\mathcal{V}}_{\mathrm{D}}$ is $J$ while the one in $\hat{\mathcal{V}}_{\mathrm{OD}}$ is $g_j$. Recall that these quantities, and also the pump strengths $\mathcal{E}_j$, are assumed to be of similar magnitude (see Fig.~2 in the main text).
By expanding the expression~\eqref{eq:bdiag1} with the help of Baker-Campbell-Hausdorf formula, one finds that the first order generator $\hat{\mathcal{G}}^{(1)}$ must satisfy
\begin{equation}
\hat{\mathcal{V}} + [\hat{\mathcal{G}}^{(1)}, \hat{\mathcal{H}}_{\mathrm{unp}}] = 0 \ .
\end{equation}
Choosing the generators to be fully off-diagonal with respect to the total excitation number,
\begin{align}
    \dbra{\bm{n} \bm{p} m}  \hat{\mathcal{G}}^{(k)}\dket{\bm{n}'\bm{p}' m'}& = 0,\quad \mbox{ for all } k \mbox{ if } \sum_j n_j = \sum_j n_j' \ ,
    \end{align}
    the first-order generator $\hat{\mathcal{G}}^{(1)}$ is then given by (for $\sum_j n_j \ne \sum_j n_j'$)
\begin{align}\label{eq:gen1}
    \dbra{\bm{n}\bm{p} m} \hat{\mathcal{G}}^{(1)}\dket{\bm{n}'\bm{p}'m'} & = \frac{ \dbra{\bm{n} \bm{p} m} \hat{\mathcal{V}}_{\mathrm{OD}}\dket{\bm{n}'\bm{p}'m'}}{\varepsilon_{\bm{n}\bm{p}m} -\varepsilon_{\bm{n}'\bm{p}'m'}}.
\end{align}
The operator $\hat{\mathcal{V}}_{\mathrm{OD}}$, given by Eqs.~\eqref{eq:Vpartsb} and \eqref{eq:HAR_m}, is a sum of terms that exchange a single excitation between the system and each cavity. Therefore, the relevant energy gaps in the denominator of Eq.~\eqref{eq:gen1}, given occupation numbers $\{\bm{n} \bm{p}\}$, will only involve $\{\bm{n}'\bm{p}'\}$ of the form $\{\bm{n}'\bm{p}'\} = \{n_0, \ldots,n_j \pm 1 \ldots, n_M, p_0, \dots, p_j\mp 1 , \ldots p_L\}$ and thus, using Eq.~\eqref{eq:upenergies},
\begin{equation}
\varepsilon_{\bm{n}\bm{p}m} -\varepsilon_{\bm{n}'\bm{p}'m'} = -\delta_j - U n_j + (m-m')\hbar\omega, \quad \mbox{ or } \quad \varepsilon_{\bm{n}\bm{p}m} -\varepsilon_{\bm{n}'\bm{p}'m'} = \delta_j + U (n_j-1) + (m-m')\hbar\omega,
\end{equation}
The perturbative expansion is thus valid as long as these gaps remain much larger than the couplings to the cavities $g_j$, for all $n_j$ and $m$, $ |\delta_j + U n_j + m\hbar\omega|\gg g_j.$ More precisely, this is true under the above-mentioned assumption that the tunnelling rate $J$ in the system and the strength $\mathcal{E}_j$ of the cavity pumps are of magnitude comparable with $g_j$, such that products like $J g_j/(\varepsilon_{\bm{n}\bm{p}m} -\varepsilon_{\bm{n}'\bm{p}'m'})$ or $\mathcal{E}_j  g_j/(\varepsilon_{\bm{n}\bm{p}m} -\varepsilon_{\bm{n}'\bm{p}'m'})$ remain small and do not make the expansion diverge. For the denominators to be sufficiently large, it is crucial that multiples $m \hbar \omega$ of Floquet driving quanta must never be close to match the bare gaps, as discussed in the main text and illustrated in Fig.~2. 
The second order generator $\hat{\mathcal{G}}^{(2)}$ is found by imposing
\begin{equation}\label{eq:G2}
[\hat{\mathcal{G}}^{(2)}, \hat{\mathcal{H}}_\mathrm{unp}]_{\mathrm{OD}} + [\hat{\mathcal{G}}^{(1)},\hat{\mathcal{V}}]_{\mathrm{OD}} + \frac{1}{2}[\hat{\mathcal{G}}^{(1)},[\hat{\mathcal{G}}^{(1)},\hat{\mathcal{H}}_\mathrm{unp}]]_{\mathrm{OD}} = 0 \ .
\end{equation}
With the given choices of $\hat{\mathcal{G}}^{(1)}$ and $\hat{\mathcal{G}}^{(2)}$, the block-diagonal operator $\hat{\mathcal{W}}$ up to second order can be expressed as~\cite{Eckardt2015}
\begin{align}\label{eq:blockdiag}
    \hat{\mathcal{W}} = & \hat{\mathcal{H}}_\mathrm{unp}+ \hat{\mathcal{V}}_{\mathrm{D}} + \frac{1}{2} [\hat{\mathcal{G}}^{(1)}, \hat{\mathcal{V}}_\mathrm{OD}]_{\mathrm{D}} \ . 
\end{align}
Therefore, what needs to be computed explicitly for determining $\hat{\mathcal{W}}$ is the first-order generator $\hat{\mathcal{G}}^{(1)}$ only.
Using the explicit expressions for $\hat{\mathcal{V}}$ from Eq.~\eqref{eq:Vparts} and the unperturbed energies of Eq.~\eqref{eq:upenergies} one finds
\begin{align}
    \dbra{\bm{n}\bm{p}m} \hat{\mathcal{V}}_{\mathrm{OD}} \dket{\bm{n}'\bm{p}'m'} =& \sum_{j=1}^L g_{j, m-m'} \Big[ \sqrt{n_j+1}\sqrt{p_j} \delta_{n_j,n_j '-1} \delta_{p_j, p_j '+1} \\
   & +(-1)^{m-m'} \sqrt{n_j}\sqrt{p_j+1} \delta_{n_j,n_j'+1} \delta_{p_j, p_j '-1}\Big]\prod_{k\ne j} \delta_{n_j,n_k'} \delta_{p_j,p_k'}.
\end{align}
Inserting into \eqref{eq:gen1}, this gives then the ``first-order'' generator
\begin{equation}
    \hat{\mathcal{G}}^{(1)} = \sum_{j=1}^L\sum_{m,m'=-\infty}^{+\infty} g_{j, m-m'}\left[ \hat{a}_j^\dagger\frac{(-1)^{m-m'} }{\delta_j + U \hat{n}_j +\hbar\omega(m-m')} \hat{c}_j -\frac{1 }{\delta_j + U \hat{n}_j -\hbar\omega(m-m')} \hat{a}_j  \hat{c}_j^\dagger\right]\ket{m}_{\! F}\!\!\bra{m'}.
\end{equation}
Returning to the non-extended space, the generator $\hat{G}^{(1)}(t)$ is time periodic and reads
\begin{equation}\label{eq:tdG1}
    \hat{G}^{(1)}(t) = \sum_{j=1}^L\sum_{m=-\infty}^{+\infty} g_{j, m}\left[ \hat{a}_j^\dagger\frac{(-1)^{m} }{\delta_j + U \hat{n}_j +\hbar\omega m} \hat{c}_j -\frac{1 }{\delta_j + U \hat{n}_j -\hbar\omega m} \hat{a}_j  \hat{c}_j^\dagger\right]e^{i m\omega t}.
\end{equation}
The fact that $\hat{G}^{(1)}(t)$ is skew-hermitian, $[\hat{G}^{(1)}(t)]^\dagger = -\hat{G}^{(1)}(t)$, can be seen by noting that $g_{j,-m}^*=(-1)^m g_{j,m}$. For completing the computation of $\hat{\mathcal{W}}$ of Eq.~\eqref{eq:blockdiag}, we further need the commutator $[\hat{\mathcal{G}}^{(1)}, \hat{\mathcal{V}}_{\mathrm{OD}}]_\mathrm{D}$. Introducing
\begin{align}
& \hat{\zeta}_{j,m}(\hat{n}_j) = \frac{1}{\delta_j + U\hat{n}_j+m\hbar\omega} \ , 
\end{align}
the commutator gives
\begin{align}\label{eq:12comm}
 \frac{1}{2} [\hat{\mathcal{G}}^{(1)}, \hat{\mathcal{V}}_{\mathrm{OD}}]_\mathrm{D} =&
\frac{1}{2}\sum_{j=1}^L \sum_{m,m',\mu=-\infty}^{+\infty}  C_{j,m-m',\mu} \Big\{
  (-1)^{m-m'} \Big(\hat{\zeta}_{j, m-m'+\mu}(\hat{n}_j-1) + \hat{\zeta}_{j,\mu}(\hat{n}_j-1) \Big)\hat{n}_j(1+\hat{c}_j^\dagger \hat{c}_j) \nonumber \\
  & -\Big(\hat{\zeta}_{j,m'-m-\mu}(\hat{n}_j) + \hat{\zeta}_{j,-\mu}(\hat{n}_j) \Big)(\hat{n}_j+1)\hat{c}_j^\dagger \hat{c}_j  \Big\} \ket{m}_{\! F}\!\!\bra{m'}\ ,
\end{align}
with $C_{j,m,\mu} = g_j^2 i^{m} e^{-i m\varphi_{j}} \mathcal{J}_{m+\mu}\left(\frac{\lambda}{\hbar\omega}\right)\mathcal{J}_{\mu}\left(\frac{\lambda}{\hbar\omega}\right).$ Terms that would involve quadratic operators of the form $\hat{c}_j^\dagger\hat{c}_j^\dagger$ and $\hat{c}_j\hat{c}_j$ are part of the off-diagonal contribution $[\hat{\mathcal{G}}^{(1)}, \hat{\mathcal{V}}_{\mathrm{OD}}]_{\mathrm{OD}}$ which is canceled by the second-order generator $\hat{\mathcal{G}}^{(2)}$ in Eq.~\eqref{eq:G2}. Terms involving two different cavities, e.g. involving operators of the form $\hat{a}_j^\dagger \hat{a}_{j'}\hat{c}_j \hat{c}_{j'}^\dagger$ with $j\ne j'$, are included in $[\hat{\mathcal{G}}^{(1)}, \hat{\mathcal{V}}_{\mathrm{OD}}]_{\mathrm{D}}$ but they ultimately vanish. The final effective Floquet Hamiltonian in the extended space can then be reconstructed inserting Eqs.~\eqref{eq:Qunp}, \eqref{eq:olVD} and \eqref{eq:12comm} into Eq.~\eqref{eq:blockdiag}. Explicitly, it reads
\begin{align}\label{eq:Wdr}
    \hat{\mathcal{W}} = & \sum_{m,m'=-\infty}^{+\infty} \Big((\hat{H}_{\mathrm{unp}} + m\hbar\omega)\delta_{m,m'} + \hat{H}_{\mathrm{tun},m-m'} \Big) \ket{m}_{\! F}\!\!\bra{m'} \nonumber  \\
    & + \frac{1}{2}\sum_{m,m'=-\infty}^{+\infty}\sum_{j=1}^L \sum_{\mu=-\infty}^{+\infty}  C_{j,m-m',\mu}\Big\{
  (-1)^{m-m'} \Big(\hat{\zeta}_{m-m'+\mu}(\hat{n}_j-1) + \hat{\zeta}_{j,\mu}(\hat{n}_j-1) \Big)\hat{n}_j \nonumber  \\
 & +\Big[(-1)^{m-m'} \Big(\hat{\zeta}_{m-m'+\mu}(\hat{n}_j-1) + \hat{\zeta}_{j,\mu}(\hat{n}_j-1) \Big)\hat{n}_j  -\Big(\hat{\zeta}_{j,m'-m-\mu}(\hat{n}_j)+ \hat{\zeta}_{j,-\mu}(\hat{n}_j) \Big)(\hat{n}_j+1) \Big]\hat{c}_j^\dagger \hat{c}_j  \Big\} \ket{m}_{\! F}\!\!\bra{m'} \ .
\end{align}
Rearranging terms and returning to the non-extended space, this leads to the time-dependent Hamiltonian 
\begin{align}\label{eq:Hdrx}
\hat{H}'(t) = \hat{H}_\mathrm{op}(t) +  \hat{H}_\mathrm{int}+\hat{H}_C + \hat{H}_{\mathrm{tun}}(t)  + \hat{H}_{\mathrm{SC}}'(t),
\end{align}
where
\begin{align}
    \hat{H}_\mathrm{op}(t) =& \frac{1}{2}\sum_{j=1}^L \sum_{m,\mu=-\infty}^{+\infty}  C_{j,m,\mu}
  (-1)^{m} \Big(\hat{\zeta}_{j,m+\mu}(\hat{n}_j-1) + \hat{\zeta}_{j,\mu}(\hat{n}_j-1) \Big)\hat{n}_j e^{i m \omega t}, \\
 \hat{H}_{\mathrm{SC}}' (t) = &\frac{1}{2} \sum_{j=1}^L \sum_{m,\mu=-\infty}^{+\infty}  C_{j,m,\mu} \Big\{
  \Big[(-1)^{m} \Big(\hat{\zeta}_{j,m+\mu}(\hat{n}_j-1) + \hat{\zeta}_{j,\mu}(\hat{n}_j-1) \Big) \nonumber \\
  & -\Big(\hat{\zeta}_{j,-m-\mu}(\hat{n}_j)+ \hat{\zeta}_{j,-\mu}(\hat{n}_j) \Big)\Big]\hat{n}_j\hat{c}_j^\dagger \hat{c}_j -\Big(\hat{\zeta}_{j,-m-\mu}(\hat{n}_j) + \zeta_{j,-\mu}(\hat{n}_j) \Big)\hat{c}_j^\dagger \hat{c}_j\Big\}  e^{i m \omega t}.
\end{align}
Additional terms describing the transformed cavity pumps are discussed below.

\subsection{High-frequency expansion}

Under the assumptions that justify the Floquet-dispersive regime described in the previous section, the dynamics can be studied separately within subspaces with different total number of excitations $N$ in the array. We now conclude the computation of a time-independent effective Hamiltonian by performing a high-frequency expansion for $\hbar\omega \gg J$. In the extended-space formalism, this corresponds to further block-diagonalizing the quasienergy operator $\hat{\mathcal{W}}$, now with respect to the Floquet index. As discussed, e.g., in Ref.~\cite{Eckardt2015}, the result of this block-diagonalization is that the Hamiltonian in the leading order is, in the non-extended space, the time averaged Hamiltonian. From the extended space representation, this can be extracted from the Floquet Hamiltonian simply by considering any diagonal block with $m=m'$ (modulo $\hbar\omega$), $\hat{H}_{\mathrm{eff}} = \bra{m} \hat{\mathcal{W}} \ket{m}_{\! F}.$
One then obtains
\begin{align}
    \hat{H}_{\mathrm{eff}} = &  \hat{H}_{\mathrm{unp}} + \hat{H}_{\mathrm{tun},0} +\sum_{j=1}^L\sum_{m=-\infty}^{+\infty} \frac{g_j^2 \mathcal{J}_m^2\left(\lambda/\hbar\omega\right)}{\delta_j + U (\hat{n}_j-1) +\hbar\omega m} \hat{n}_j + \nonumber \\
    & \sum_{j=1}^L \sum_{m=-\infty}^{+\infty} g_j^2  \mathcal{J}_m^2\left(\frac{\lambda}{\hbar\omega}\right) \Big[ \frac{1}{\delta_j + U (\hat{n}_j-1) +\hbar\omega m}\hat{n}_j -\frac{1}{\delta_j + U \hat{n}_j -\hbar\omega m}(\hat{n}_j+1) \Big]\hat{c}_j^\dagger \hat{c}_j , \nonumber
\end{align}
which can be rewritten in the form of Eq. (3) of the main text,
\begin{equation}\label{eq:Heffapp}
 \hat{H}_{\mathrm{eff}}  =   \hat{H}_{\mathrm{unp}}  - \sum_{\langle j, j'\rangle} \Big(J^{\mathrm{eff}}_{j j'} e^{i\theta^{\mathrm{eff}}_{jj'}} \hat{a}_{j'}^\dagger \hat{a}_{j} + \mathrm{h.c.}\Big) + \sum_{j=1}^L \Big(\xi_j(\hat{n}_j-1)\hat{n}_j +\chi_j(\hat{n}_j)\hat{c}_j^\dagger \hat{c}_j\Big) , 
\end{equation}
with the functions $\xi_j(\hat{n}_j)$ and $\chi_j(\hat{n}_j)$ given in Eq. (4), namely
\begin{align}
& \xi_j(\hat{n}_j) = \sum_{m=-\infty}^{+\infty} \frac{g_j^2 \mathcal{J}_m^2\left(\lambda/\hbar\omega\right)}{\delta_j + U (\hat{n}_j-1) +\hbar\omega m} ,\\
& \chi_j(\hat{n}_j) = \sum_{m=-\infty}^{+\infty} g_j^2 \mathcal{J}_m^2\left(\frac{\lambda}{\hbar\omega}\right) \Big[ \frac{U}{(\delta_j + U (\hat{n}_j-1) -\hbar\omega m)(\delta_j + U \hat{n}_j -\hbar\omega m)} \hat{n}_j -\frac{1}{\delta_j + U \hat{n}_j -\hbar\omega m}\Big].
\end{align}
In deriving the last expression, we have used the fact that $\mathcal{J}_m^2\left(\lambda/\hbar\omega\right) = \mathcal{J}_{-m}^2\left(\lambda/\hbar\omega\right)$ to change $m\to -m$ in one of the terms. The Hamiltonian $\hat{H}_{\mathrm{eff}}$ is then the result of our perturbative treatment for the coherent dynamics. 

For potential interest in the study of the single excitation dynamics, we restrict the effective Hamiltonian of Eq.~\eqref{eq:Heffapp} to the single-excitation subspace, spanned by the states $\hat{a}_j^\dagger\ket{0}$. For the terms $\xi_j(\hat{n}_j-1)\hat{n}_j$ and $\chi_j(\hat{n}_j)$, this simply amounts to substituting $\hat{n}_j\to 1$ in the denominators. The term $\xi_j(\hat{n}_j)$ gives instead
\begin{align}
 \sum_{\ell,\ell'}\sum_{j=1}^L \hat{a}^\dagger_\ell\ket{0}\bra{0}\hat{a}_\ell   \frac{1}{\delta_j + U \hat{n}_j -\hbar\omega m} \hat{a}^\dagger_{\ell'}\ket{0}\bra{0}\hat{a}_{\ell'} =&  
 \sum_{j=1}^L \frac{1}{\delta_j -\hbar\omega m} \mathbb{1}_\mathrm{1p} + \sum_{j=1}^L \frac{-U}{(\delta_j + U -\hbar\omega m)(\delta_j -\hbar\omega m)}\hat{n}_{j} \ ,  
\end{align}
where $\mathbb{1}_\mathrm{1p} = \sum_\ell \hat{a}_\ell^\dagger \ket{0}\bra{0}\hat{a}_\ell$ is the identity operator in the single excitation subspace. 
The single-excitation effective Hamiltonian $  \hat{H}_{\mathrm{eff}}^{(1)}$ is thus
\begin{equation}
  \hat{H}_{\mathrm{eff}}^{(1)} = \hat{H}_{\mathrm{unp}} + \hat{H}_{\mathrm{tun},0} +\sum_{j=1}^L\Big[ \xi_j^{(1)} (\hat{n}_j -\hat{c}_j^\dagger \hat{c}_j)+ \chi_j^{(1)}\hat{n}_j\hat{c}_j^\dagger \hat{c}_j \Big] \ , \label{eq:Heffsm}
\end{equation}
with
\begin{equation}
 \xi_j^{(1)} = \sum_{m=-\infty}^{+\infty}  g_j^2 \frac{\mathcal{J}_m^2\left(\lambda/\hbar\omega\right)}{\delta_j -\hbar\omega m}, \qquad
\chi_j^{(1)} = \sum_{m=-\infty}^{+\infty}  g_j^2 \frac{2U \mathcal{J}_m^2(\lambda/\hbar\omega)}{(\delta_j  -\hbar\omega m)(\delta_j + U  -\hbar\omega m)} \ .
\end{equation}

\subsection{Cavity pumps} 
We did not yet consider the impact of the block-diagonalization with respect to $N$ on the cavity pump term, such that block-diagonalizing the array-cavity coupling with respect to $N$ was equivalent to block-diagonalizing the whole quasienergy operator. In order to introduce the cavity pumps at the level of the effective Hamiltonian, one needs to compute how these terms are modified by the perturbative transformation. In summary, the transformation introduces off-diagonal driving terms for the artificial atoms that couple to a cavity. These terms are largely off-resonant and can thus be neglected in our applications. 

Focusing on each cavity pump at a time, it is convenient to work in a frame rotating at the corresponding pump frequency $\omega_j$ for the whole system (similarly to how one would approach a multilevel system subject to multiple pulses, where the effect of each pulse can be better understood in a frame rotating at the corresponding drive frequency). In this frame, the pump term becomes time independent, 
\begin{equation}
    \hat{H}_{j,\mathrm{pump}} = \mathcal{E}_j\hat{c}^\dagger_j  + \mathcal{E}_j^* \hat{c}_j  \ .
    \end{equation}
    and the atoms and cavities energies acquire a potential shift $-\hbar\omega_j$. We can then perform the Floquet-dispersive regime transformation and high-frequency expansion for the $j$th cavity in this frame. The resulting effective Hamiltonian terms are the same as in Eq.~\eqref{eq:Heffapp}, with the additional potential shift $-\hbar\omega_j$ for atoms and cavities.

The Floquet-dispersive regime transformation $U(t)$ then acts on the pump according to 
\begin{align}\label{eq:Hp_U}
    \hat{U}^\dagger(t) \hat{H}_{j,\mathrm{pump}} \hat{U}(t) = \sum_{j=1}^L [\mathcal{E}_j  \hat{U}^\dagger(t) \hat{c}^\dagger_j \hat{U}(t)+ \mathcal{E}_j^* U^\dagger(t) \hat{c}_j \hat{U}(t) ] \ . 
\end{align}
The transformed cavity operators can be computed perturbatively through Hausdorf expansion, finding
\begin{subequations}\label{eq:cjt}
\begin{align}
     \hat{c}_j(t) \equiv U^\dagger(t) \hat{c}_j U(t) \simeq \hat{c}_j + \sum_{m=-\infty}^{+\infty} \frac{g_{j,m}}{\delta_j+U \hat{n}_j -m\hbar\omega}\hat{a}_j e^{i m \omega t},\\
     \hat{c}_j^\dagger(t) \equiv U^\dagger(t) \hat{c}^\dagger_j U(t) \simeq \hat{c}_j^\dagger +  \sum_{m=-\infty}^{+\infty} \hat{a}^\dagger_j \frac{g^*_{j,-m}}{\delta_j+U \hat{n}_j +m\hbar\omega} e^{i m \omega t},
\end{align}
\end{subequations}
with $g_{j,m}$ given in Eq.~\eqref{eq:Jjmgjm}. Note that we have kept only the first commutator term in the Hausdorf expansion since $\mathcal{E}_j\sim g_j$ in our model, and thus the single-commutator term is already of second order in the perturbative expansion when inserted into Eq.~\eqref{eq:Hp_U}.
The leading order in the high-frequency expansion is then the time average of these operators, namely
\begin{align}
     \hat{c}_{j,0} = \hat{c}_{j} + \frac{g_j \mathcal{J}_0(\lambda/\hbar\omega)}{\delta_j+U \hat{n}_j}\hat{a}_j,\qquad (\hat{c}_j^\dagger)_0 =  \hat{c}_{j}^\dagger  + \hat{a}^\dagger_j \frac{g_j \mathcal{J}_0(\lambda/\hbar\omega)}{\delta_j+U \hat{n}_j}.
\end{align}
The corrections to $\hat{c}_j$ and $\hat{c}_j^\dagger$ thus describe an extra drive of the artificial atoms connecting manifolds with different excitation number. The whole transformed pump Hamiltonian then becomes
\begin{equation}
\hat{H}_{j,\mathrm{pump}}'= \hat{H}_{j,\mathrm{pump}} + \sum_{j=1}^L \Big(\frac{\widetilde{\mathcal{E}}_j }{\delta_j + U \hat{n}_j}\hat{a}_j+ \hat{a}_j^\dagger\frac{ \widetilde{\mathcal{E}}_j^*}{\delta_j + U \hat{n}_j }\Big),
\end{equation}
with $\widetilde{\mathcal{E}}_j = g_j \mathcal{E}_j\mathcal{J}_0(\lambda/\hbar\omega)$.
In the undriven system, this effective driving Hamiltonian for the artificial atoms is used for exciting the array (implementing single-qubit rotations in quantum information processing applications) if the pump frequency is resonant with the atomic transitions. Here, since the cavity pumps are far detuned and under the assumptions of validity of the Floquet-dispersive regime and high-frequency expansion, these extra driving terms on the array are far off-resonant and can be neglected. Returning to the frame rotating at $\Delta/\hbar$, the cavity pump term in conclusion remains unaltered,
\begin{equation}
\hat{H}_{j,\mathrm{pump}}^{\mathrm{eff}}(t)= \hat{H}_{j,\mathrm{pump}}(t) = 
     \mathcal{E}_j\hat{c}^\dagger_j e^{-i\omega_j t} + \mathcal{E}_j^* \hat{c}_j e^{i\omega_j t} \ .
\end{equation}

\subsection{Effective dissipator}
The perturbation theory discussed for the Hamiltonian can be reformulated at the superoperator level to compute an effective dissipator via the same procedure. The master equation~\eqref{eq:mastereq} in the frame where the Floquet drive has been integrated out can be rewritten as 
$d\hat{\rho}/dt = \mathcal{L}(t) \hat{\rho}$, with Lindbladian superoperator
\begin{equation}
\mathcal{L}(t) = -\frac{i}{\hbar}[\hat{H}(t),\cdot] + \sum_j \kappa_j \mathcal{D}[\hat{c}_j](\cdot),
\end{equation}
and Hamiltonian $\hat{H}(t)$ given in Eq.~\eqref{eq:wideH}. 
The unitary Floquet-dispersive regime transformation $\hat{U}(t)$ that block-diagonalizes the Hamiltonian with respect to the total number of atom excitations transforms the Lindbladian $\mathcal{L}(t)$ into 
\begin{equation}
\mathcal{L}'(t) = \hat{U}^\dagger(t) \mathcal{L}(t) \hat{U}(t) = \frac{i}{\hbar}[\hat{H}'(t),\cdot]  +  \sum_j \kappa_j  \mathcal{D}[\hat{U}(t)^\dagger \hat{c}_j \hat{U}(t)](\cdot).
\end{equation}
with $\hat{H}'(t) = \hat{U}^\dagger(t)\hat{H}(t)\hat{U}(t) - i \hbar \hat{U}^\dagger(t)\partial_t \hat{U}(t)$.
To second order, the transformed Hamiltonian $\hat{H}'(t)$ is the Floquet-dispersive-regime Hamiltonian given explicitly in Eq.~\eqref{eq:Hdrx}. 
Focusing on the dissipative part, one can expand the operators $\hat{c}_j(t) = \hat{U}^\dagger(t)\hat{c}_j \hat{U}(t)$ as a Fourier series,
\begin{align}
    & \hat{c}_j(t) = \sum_{m=-\infty}^{+\infty}\hat{c}_{j,m} e^{i m \omega t} ,\\
    & \hat{c}^\dagger_j(t) = \sum_{m=-\infty}^{+\infty}(\hat{c}_{j,m})^\dagger e^{-i m \omega t} .
\end{align}
The leading-order expressions for $\hat{c}_{j,m}$ are found straightforwardly from Eq.~\eqref{eq:cjt}, leading to the expression given in the main text, Eq. (5). The dissipators then read
\begin{equation}
  \mathcal{D}[\hat{c}_{j}(t)](\cdot) = \sum_{m,m'=-\infty}^{+\infty}\Big\{ \hat{c}_{j,m}(\cdot) [\hat{c}_{j,m'}]^\dagger -\frac{1}{2}[\hat{c}_{j,m'}]^\dagger \hat{c}_{j,m}(\cdot)-\frac{1}{2}(\cdot) [\hat{c}_{j,m'}]^\dagger \hat{c}_{j,m}\Big\} e^{i(m-m')\omega t}.
\end{equation}
Redefining $m'=m-\mu$, this becomes
\begin{equation}
  \mathcal{D}[\hat{c}_{j}(t)](\cdot) = \sum_{m,\mu=-\infty}^{+\infty}\Big\{ \hat{c}_{j,m}(\cdot) [\hat{c}_{j,m-\mu}]^\dagger -\frac{1}{2}[\hat{c}_{j,m-\mu}]^\dagger \hat{c}_{j,m}(\cdot)-\frac{1}{2}(\cdot) [\hat{c}_{j,m-\mu}]^\dagger \hat{c}_{j,m}\Big\} e^{i \mu \omega t},
\end{equation}
such that the $\mu$-th Fourier component of the dissipator can be immediately identified.
Finally, the time average of the transformed Lindbladian following from the high-frequency expansion selects the zero-frequency component only, which is
\begin{equation}\label{eq:Deff1}
      \mathcal{D}_{\mathrm{eff}}(\cdot) = \sum_{j=1}^L \kappa_j \sum_{m=-\infty}^{+\infty}\Big\{ \hat{c}_{j,m}(\cdot) [\hat{c}_{j,m}]^\dagger -\frac{1}{2}[\hat{c}_{j,m}]^\dagger \hat{c}_{j,m}(\cdot)-\frac{1}{2}(\cdot) [\hat{c}_{j,m}]^\dagger \hat{c}_{j,m}\Big\}.
\end{equation}
The time averaged effective Lindbladian then reads
\begin{equation}\label{eq:me_eff}
\mathcal{L}_{\mathrm{eff}} = -\frac{i}{\hbar}[\hat{H}_\mathrm{eff}, \cdot] +\mathcal{D}_\mathrm{eff}(\cdot),
\end{equation}
with the effective Hamiltonian of Eq.~\eqref{eq:Heffapp}.
Simulations of the effective dynamics are performed using this master equation, including the cavity pumps.

\section{Reservoir engineering for the stroboscopic dynamics}
\label{sec:reservoir_engineering}

 We now discuss how reservoir engineering is realized at the level of the effective stroboscopic Hamiltonian of the array shaped through the Floquet drives. Our starting point is the effective master equation~\eqref{eq:me_eff} derived in the previous Section, neglecting the effective excitation decay from the array (see section ``Postselection''). It reads
\begin{equation}\label{eq:arrcavme}
    \frac{d\hat{\rho}}{dt} =- \frac{i}{\hbar}[\hat{H}_\mathrm{eff}, \hat{\rho}] + \sum_j \kappa_j  \mathcal{D}[\hat{c}_{j}](\hat{\rho}).
\end{equation}
The theory developed here follows similar ones done in other contexts~\cite{Murch2012, Hacohen2015}, with the difference that here multiple cavities are considered and that the relevant system Hamiltonian is the Floquet-engineered Hamiltonian with artificial magnetic flux. The effective Hamiltonian can be rewritten in the form
\begin{equation}\label{eq:effboh}
 \hat{H}_{\mathrm{eff}} = \hat{H}_{\mathrm{S}}^{\mathrm{eff}}  + \sum_{j=1}^L \Big[\chi_j (\hat{n}_j) \hat{c}_j^\dagger \hat{c}_j + \delta_j \hat{c}_j^\dagger \hat{c}_j  + \mathcal{E}_j \hat{c}_j^\dagger e^{-i \omega_j t}+\mathcal{E}_j^* \hat{c}_j e^{i \omega_j t}\Big],
\end{equation}
where the effective atomic Hamiltonian is 
\begin{equation}
\hat{H}_\mathrm{S}^{\mathrm{eff}}  = \sum_{j=1}^M \Big\{ \xi_j(\hat{n}_j-1) \hat{n}_j + \frac{U}{2}\hat{n}_j(\hat{n}_j-1)\Big\}- \sum_{\langle j,j'\rangle} J^{\mathrm{eff}}_{jj'}(e^{i\theta^\mathrm{eff}_{jj'}}\hat{a}_j\hat{a}_{j'}^\dagger+\mathrm{h.c.}).
\end{equation}
Note that we have included the effective term $\sum_j \xi_j(\hat{n}_j-1)\hat{n}_j$ into the effective atomic Hamiltonian, slightly differently from Eq.~(3) of the main text. Since the effective atom-cavity coupling is diagonal with respect to the cavity Fock basis, we can make a gauge transformation for the cavities which eliminates the time-dependence given by the cavity pump terms (corresponding to a frame rotating at $\omega_j$ for the $j$th cavity) without altering the atomic part. The resulting $\hat{H}_{\mathrm{eff}}$ reads
\begin{equation}
 \hat{H}_{\mathrm{eff}} = \hat{H}_{\mathrm{S}}^{\mathrm{eff}}  + \sum_{j=1}^L \Big[\chi_j (\hat{n}_j) \hat{c}_j^\dagger \hat{c}_j + d_j \hat{c}_j^\dagger \hat{c}_j  + \mathcal{E}_j \hat{c}_j^\dagger +\mathcal{E}_j^* \hat{c}_j \Big], 
\end{equation}
where we have introduced the cavity-pump detunings $d_j=\delta_j - \hbar\omega_j$. 
We displace the cavity fields in order to remove the terms describing the cavity pumps. The displacement is described by the operators $\hat{D}_j(\alpha_j) = e^{\alpha_j \hat{c}_j^\dagger - \alpha_j^* \hat{c}_j}$ such that $\hat{D}_j^\dagger(\alpha_j) \hat{c}_j \hat{D}_j(\alpha_j) = \hat{c}_j + \alpha_j.$ In the absence of coupling to the array, the displaced cavity modes would look like damped (but not pumped) harmonic oscillators relaxing to the vacuum. This vacuum corresponds, for the non-displaced fields, to coherent states $\ket{\alpha_j}=\hat{D}(\alpha_j)\ket{0}$ having mean photon number $\bar{n}_{\mathrm{ph},j}=|\alpha_j|^2$. 
The field displacement changes the Hamiltonian~\eqref{eq:effboh} according to
\begin{align}
  \hat{H}_{\mathrm{eff}} = \hat{H}_{\mathrm{S}}^{\mathrm{eff}} + \sum_{j=1}^L \Big\{ \chi_j (\hat{n}_j) \big[\hat{c}_j^\dagger \hat{c}_j +\alpha_j^* \hat{c}_j +\alpha_j \hat{c}_j^\dagger +|\alpha_j|^2 \big] + d_j \big[\hat{c}_j^\dagger \hat{c}_j +\alpha_j^* \hat{c}_j +\alpha_j \hat{c}_j^\dagger +|\alpha_j|^2 \big] +  \mathcal{E}_j \hat{c}_j^\dagger +\mathcal{E}_j^* \hat{c}_j \Big\},
  \label{eq:Halpha}
\end{align}
where we have dropped constant energy shifts. Moreover, it changes the dissipative part of the master equation as follows,
\begin{align}
   \kappa_j  \hat{D}_j^\dagger(\alpha_j) \mathcal{D}[\hat{c}_j]\hat{\rho} \hat{D}_j(\alpha_j) = -\frac{i}{\hbar}\Big[\hat{H}_{\alpha_j} ,\hat{\rho} \Big] + \kappa_j \mathcal{D}[\hat{c}_j]\hat{\rho}, \qquad \hat{H}_{\alpha_j} = \frac{i \hbar \kappa_j}{2}(\alpha_j^* \hat{c}_j - \alpha_j \hat{c}_j^\dagger).
\end{align}
The amplitudes $\alpha_j$ are chosen such as to cancel all terms proportional to $\hat{c}_j$ and $\hat{c}^\dagger_j$ that do not couple to the system in Eq.~\eqref{eq:Halpha}, which yields 
\begin{equation}
    \alpha_j = -\frac{\mathcal{E}_j}{d_j - i\hbar \kappa_j/2 }.
\end{equation}
The master equation for the displaced density matrix is then of the form of Eq.~\eqref{eq:arrcavme} with displaced effective Hamiltonian
\begin{equation}\label{eq:Heffalpha}
    \hat{H}_{\mathrm{eff}} = \hat{H}_{\mathrm{S}}^{\mathrm{eff}} + \sum_{j=1}^L \Big\{ \chi_j (\hat{n}_j) \big[\hat{c}_j^\dagger \hat{c}_j + \alpha_j^* \hat{c}_j + \alpha_j \hat{c}_j^\dagger + |\alpha_j|^2\big] +  d_j \hat{c}_j^\dagger \hat{c}_j \Big\}.
\end{equation}
 Let us recall that the Hamiltonian $\hat{H}_{\mathrm{S}}^{\mathrm{eff}}$  is the atomic Hamiltonian that is Floquet engineered to feature artificial magnetic flux and whose eigenstates we wish to prepare and stabilize. We represent the Hamiltonian of Eq.~\eqref{eq:Heffalpha} in the basis of the eigenstates of $
  \hat{H}_{\mathrm{S}}^{\mathrm{eff}}  = \sum_{\mu} \varepsilon_\mu \ket{\mu}\!\bra{\mu}, 
$
obtaining
\begin{equation}
    \hat{H}_{\mathrm{eff}} = \sum_{\mu} \varepsilon_\mu \ket{\mu}\!\bra{\mu} + \sum_{j=1}^L \sum_{\mu,\mu'}\Big\{ \bra{\mu}\chi_j(\hat{n}_j)\ket{{\mu'}} \ket{\mu}\! \bra{{\mu'}} \Big(\hat{c}_j^\dagger \hat{c}_j + \alpha^* \hat{c}_j + \alpha_j \hat{c}_j^\dagger + |\alpha_j|^2 \Big) +  d_j \hat{c}_j^\dagger \hat{c}_j \Big\}.
\end{equation}
In interaction picture with respect to the first and last term, this further becomes
\begin{equation}
    \hat{H}_{\mathrm{eff}}(t) =  \sum_{j=1}^L \sum_{\mu,\mu'} \bra{\mu}\chi_j(\hat{n}_j)\ket{{\mu'}} \ket{\mu}\! \bra{{\mu'}} \Big[(\hat{c}_j^\dagger \hat{c}_j + |\alpha_j|^2) e^{i (\varepsilon_{\mu}-\varepsilon_{\mu'}) t} + \alpha_j^* \hat{c}_j e^{i ( \varepsilon_{\mu}-\varepsilon_{\mu'}- d_j) t} +\alpha_j \hat{c}_j^\dagger e^{i (\varepsilon_{\mu}-\varepsilon_{\mu'} + d_j) t} \Big] .
\end{equation}
From this expression one can see that by choosing $d_j = \pm(\varepsilon_\mu-\varepsilon_{\mu'}) $, so by setting the detuning of the pump from the $j$-th cavity resonance to match the effective transition energy $\varepsilon_\mu - \varepsilon_{\mu'}$ of the system, terms of the form $ \ket{\mu}\! \bra{{\mu'}}  \hat{c}_j, \, \ket{{\mu}}\! \bra{{\mu'}}  \hat{c}_j^\dagger, $ become resonant. These terms describe system transitions between the effective eigenstates $\ket{{\mu'}}$ and $\ket{{\mu}}$ which are accompanied by the absorption or emission of a photon (in the displaced modes) in the $j$-th cavity. 
For the other terms in the Hamiltonian not to compete with such processes and thus being negligible in rotating-wave-approximation, it must hold that the system effective energy gaps are much larger than the effective atom-cavity coupling,
\begin{equation}
    |\varepsilon_\mu -\varepsilon_{\mu'}| \gg |\!\bra{\mu} \chi_j(\hat{n}_j) \ket{{\mu'}}\!|\sqrt{n_{\mathrm{ph},j}} \ .
\end{equation}
Also, it must hold that the difference between the gaps of the resonant and non-resonant processes are much larger than the coupling,
\begin{equation}
  |(\varepsilon_\mu -\varepsilon_{\mu'}) -  (\varepsilon_\eta -\varepsilon_{\eta'})|\gg |\!\bra{\mu} \chi_j(\hat{n}_j) \ket{{\mu'}}\!| \sqrt{n_{\mathrm{ph},j}} ,
\end{equation}
where we have fixed $d_j=\varepsilon_\eta - \varepsilon_{\eta'}$ for some $\eta$. The latter condition is needed if one really aims at addressing a single transition. However, in the applications studied the fact that multiple effective transitions have a similar gap is advantageous, since they can all be controlled with a single cavity. Under the above conditions, we can focus on a single resonant transition, for clarity, and the cavity used to control it. The Hamiltonian then reduces to
\begin{equation}
    \hat{H}_{\eta\eta'j} = \overline{\chi}_{\eta \eta'j}\ket{{\eta}}\!\bra{{\eta'}}  \hat{c}_j^\dagger +\overline{\chi}^*_{\eta'\eta j} \ket{{\eta'}}\!\bra{{\eta}} \hat{c}_j  \ ,
\end{equation}
where we have defined the rate $\overline{\chi}_{\eta\eta'j}= \alpha_j \bra{\eta} \chi_j (\hat{n}_j) \ket{{\eta'}}$. In this reduced model, the master equation now describes a Jaynes-Cummings interaction between the two-level system comprising $\ket{\varepsilon_{\eta}}$ and $\ket{\varepsilon_{\eta'}}$ and the cavity with photon leakage from the cavity,
\begin{equation}
  \frac{d\hat{\rho}}{dt} =  -\frac{i}{\hbar}[\hat{H}_{\eta\eta'j}, \hat{\rho}] +\kappa_j \mathcal{D}[\hat{c}_j](\hat{\rho})\ .
\end{equation}
This equation can be solved in the ``bad-cavity'' limit, $\hbar \kappa_j \gg \overline{\chi}_{\eta\eta'j}$ ~\cite{ScullyZubairy1997}. The solution features a dissipative transition from $\ket{{\eta}'}$ to $\ket{{\eta}}$ in which population is transferred exponentially fast, $p_{\eta'}(t) = e^{-\Gamma_{\eta'\to\eta}^{(j)}t}p_{\eta'}(0)$, with rate
\begin{align}\label{eq:jcrate}
    \Gamma_{\eta'\to\eta}^{(j)} = & 4\frac{\bar{n}_{\mathrm{ph},j}}{\hbar^2\kappa_j}\ \lvert\bra{{\eta}}\chi_j(\hat{n}_j) \ket{{\eta}'}\rvert^2. 
\end{align}
The transition takes place also if the system departs from the bad-cavity limit ($\hbar\kappa_j\lesssim \overline{\chi}_{\eta\eta'j}$), but with more complex transient dynamics~\cite{Murch2012}. This is due to the fact that, in this case, the atom-cavity system enters an effective strong-coupling regime and can exhibit damped coherent evolution, rather than purely monotonic decay. In conclusion, each cavity can be exploited to induce a dissipative transition at rate $\Gamma_{\eta'\to\eta}^{(j)}$ between two eigenstates (more generally, two energy resolvable subspaces) of the effective atomic Hamiltonian obtained through Floquet engineering. In some of the examples shown in the main text, a single cavity is used to control multiple transitions. This is possible, and can be exploited, whenever the corresponding transition energies differ by less than the cavity decay energy $\hbar\kappa$. Indeed, the rate for a generic, non-resonant, transition can be estimated via Fermi's golden rule~\cite{Hacohen2015}, and is proportional to the cavity spectral density. For a pumped-damped cavity, the latter has Lorentzian shape of width $\kappa$ centered at the cavity-pump detuning~\cite{Clerk2010, Murch2012, Hacohen2015}. Hence, if a transition is activated resonantly by a cavity, other transitions with an energy gap in a nearby window of width $\hbar\kappa$ will be activated with a sizeable rate, since they still fall within the width of the Lorentzian peak. 

 \section{Numerical simulations, postselection, and further examples}\label{app:numerical_simulations}

\subsection{Numerical simulations} \label{sec:numerical_simulations}

Numerical simulations of the full driven model and of the effective model are done using the master equations \eqref{eq:mastereq} and \eqref{eq:me_eff}, respectively, in both cases after having performed an additional transformation. Indeed, in order to reduce the number of cavity states to be included in the truncated Hilbert space, a displacement of all the cavity modes is first performed to eliminate linear terms in the cavity operators (not coupling to the array) from the master equation, following the same steps done in Sec.~\ref{sec:reservoir_engineering}. When pumped, each cavity initially relaxes to the coherent state described by the displacement operator, corresponding to the vacuum of the displaced fields, over a timescale much faster than that characterizing the dynamics we want to study. Therefore, in the simulations each cavity is initialized directly in the vacuum of the displaced field and states with up to three photons are included per cavity. For the array, the driven dynamics with $N$ total excitations is simulated including states with up to $N+1$ total excitations, since there are no non-zero matrix elements associated with the simultaneous creation or loss of more than a single extra excitation. Only states up to $N$ excitations instead are kept for the effective master equation, since higher excitation states are completely decoupled. All the Floquet-dressed coefficients entering the effective Hamiltonian, Eq. (4) in the main text, are computed with the truncation $m\in[-50,50]$. The same truncation applies to the effective jump operators, Eq. (5). In all simulations expect for those reported in Fig.~3(d) (concerning the two-excitation example), the master equation is directly integrated numerically in time. The results of Fig.~3d are computationally more demanding, and are obtained instead by using the Monte Carlo wavefunction method~\cite{Molmer1993} and averaging over 350 trajectories. The code for solving the master equations in time makes extensive use of the QuTip Python library~\cite{Qutip1, Qutip2}.

\subsection{Postselection}

We here discuss further how it is possible, in practice, to probe and verify interesting properties characterizing the states prepared, despite the weak excitation loss due to Purcell decay that could in principle spoil the quantum simulation, by simply postselecting on the measurement results. % 
In general, one can postselect by measuring observables that carry also information about the total number of excitations, such that a record can be discarded if the latter was not conserved. The site densities and currents that we present in Figs. 3 and 4 belong indeed to this class, and are at the same time key signatures characterizing the states stabilized. Site densities are measured directly through standard dispersive readout of all atoms, from which the total number of excitations is also determined allowing postselection. These quantities immediately allow one to observe the typical density patterns of flux ladders, which are more generally an important signature of quantum-Hall-like states. Also, they allow observation of the signature interference pattern in the Aharonov-Bohm cage example, Sec.~\ref{sec:ABcage}. Currents can be measured using the method proposed in Ref. \cite{Kessler2014} and experimentally realized in cold atoms, e.g., in \cite{Atala2014}. This involves biasing the atom frequencies (which is the control knob we use to perform Floquet engineering) before measuring the excitation number of each atom via dispersive readout. More in detail, one performs a sequence of measurements in which pairs of neighbouring atoms are first detuned from other atoms, suppressing tunneling between different pairs, and then site occupations are measured. The time evolution of the latter, obtained by measuring after different idle times, will exhibit oscillations which are directly determined by the current between the two sites, allowing its estimation~\cite{Kessler2014}. Since the frequency biasing does not change the atomic excitations, and the measurements also give the total excitation number, postselection can be performed while currents are measured.

In numerical simulations, where we can access the full density matrix $\hat{\rho}$, postselection in reproduced by projecting the density matrix in the subspace of conserved excitations (defined by the projector $\hat{P}$) when computing observables, $\rho_{\mathrm{ps}}=\hat{P}\rho\hat{P}/\mathrm{tr}[\hat{P} \hat{\rho}]$, which identifies the discarded fraction as $p = 1-\mathrm{tr}[\hat{P}\hat{\rho}]$. In Fig. 3 and 4 we show the occupation of the effective eigenstates computed in this way, despite being hardly measurable directly, in order to unambiguously certify the success of the protocols, in addition to showing measurable observables such as site occupations and currents.

\begin{figure}
\includegraphics[width=\textwidth]{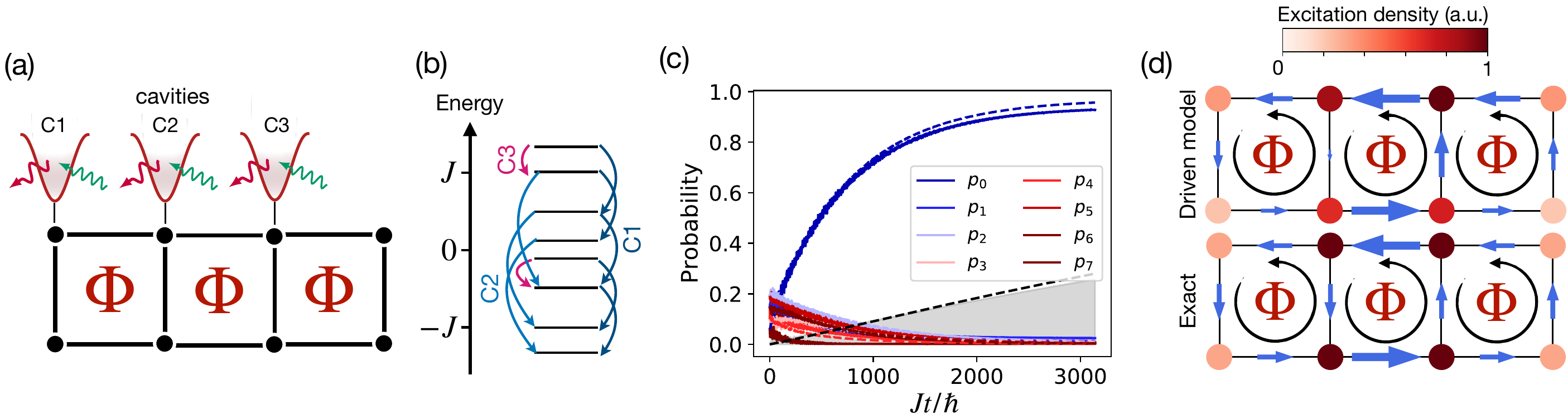}
\caption{Ground state preparation for an excitation in a three-plaquette ladder. (a) System geometry and cavity couplings. (b) Effective energy spectrum and transitions induced via reservoir engineering. (c) Build up of population in the effective ground state (solid, for the full driven model, dashed, for the effective master equation) induced by the Floquet-dissipative scheme. The grey shaded area indicates the fraction of discarded data in postselection in the full model, with the corresponding prediction given by the effective master equation (black dashed) (d) Final current patterns (the size of the arrows reproduces the current strength in arbitrary units) and site densities (color scale) in the final state of panel (c) [upper panel] as compared to the ideal ground state [lower panel]. The parameters used are $\Phi=0.4 \pi$, $\hbar\omega=20J$, $\lambda=0.72\omega$, $\bm{\delta}/\hbar\omega = (1.74 ,1.71, 1.764)$, $U=8J$, $\bm{\kappa}/J = (0.09, 0.09,0.09)$, $\bm{\mathcal{E}}/J=(1,1.7,0.4)$, $\bm{g}/J =(1,1,1)$.}
\label{fig:3plaq}
\end{figure}

\subsection{More on the stabilization in the bosonic ladder}

Considering the ladder system of Fig. 3 of the main text, we label the sites with indices $(\ell,r)$ indicating the leg ($\ell=1$ for the upper leg, $\ell=2$ for the lower), and the rung, $r=1,2,3$. The current on the edges along the legs ($j_{\ell;r,r+1}^{\paral}$) and along the rungs ($j_{r}^{\perp}$) are computed as the expectation value of the operators~\cite{Hugel2014, Piraud2015, Atala2014}
\begin{subequations}
\begin{align}
& \hat{j}_{\ell;r,r+1}^{\paral} = i J (\hat{a}_{\ell,r+1}^\dagger \hat{a}_{\ell,r} -\hat{a}_{\ell,r+1} \hat{a}_{\ell,r}^\dagger) ,\\
& \hat{j}_{r}^{\perp} = i J (e^{-i r \Phi} \hat{a}_{1,r}^\dagger \hat{a}_{2,r} -e^{i r \Phi}\hat{a}_{1,r} \hat{a}_{2,r}^\dagger) .
\end{align}
\end{subequations}

A further example of ground state preparation in a small bosonic ladder is reported in Fig.~\ref{fig:3plaq}. In this case, a single excitation in a three-plaquettes is considered, coupled to three cavities. A final current pattern in agreement with the ideal one is achieved. Better agreement is obtained at longer stabilization times.

\subsection{Preparing Aharonov-Bohm cages} \label{sec:ABcage}

\begin{figure}
\centering
\includegraphics[width=0.8\linewidth]{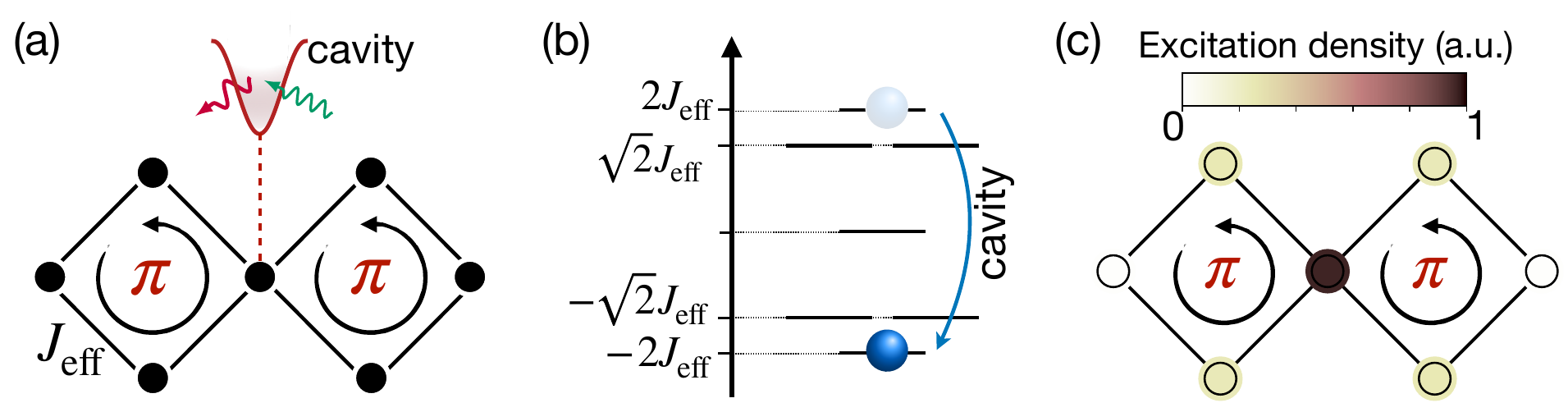}
\caption{(a) Rhombic geometry and (b) single-excitation energy spectrum of the effective array Hamiltonian. The cavity couples at the central site and cools the state component in the most excited state to the ground state. (c) Final localized site density exhibiting Aharonov-Bohm caging. The parameters used are $\hbar\omega=25J$, $\delta_{1} = 34.8J$, $U=12J$, $\kappa_1=0.2J/\hbar$, $\mathcal{E}_1=2.7J$, $g_1=J$, final time $Jt/\hbar=600$.}
\label{figAB}
\end{figure}

We here consider the diamond-chain of corner-sharing rhombic 
plaquettes with flux $\Phi=\pi$ [Fig.~\ref{figAB}(a)]. It exhibits so called Aharonov Bohm cages, single-excitation states that are strictly localized on each x-shaped group of five sites via destructive interference~\cite{Vidal2000,Mukherjee2018, Kremer2020}. While these states form a flat Bloch band in an extended chain, one of them exists already as the ground state of a minimal model with two plaquettes. It can be prepared in a controlled fashion by starting from a single excitation on the central site (having equal overlap with both the most excited and the ground state) using a single resonator, as shown in Fig.~\ref{figAB}(b). The site occupations after the preparation are depicted in Fig.~\ref{figAB}(c), confirming the achievement of the desired effect. In a large chain this preparation scheme can easily be scaled up to prepare localized excitations at targeted positions.

\section{Choosing the parameters and an illustrative example} \label{sec:example}

In this section we discuss further and exemplify how the parameters for the driven-dissipative stabilization protocols are chosen. 

\subsection{General conditions on parameters}

The values of the fundamental system parameters are chosen to be typical of state-of-the-art circuit QED platforms. In particular, we probed values such as $U \sim 8-15J$, $g_j=J$, $\kappa_j \sim 0.05-0.2J$, $\delta_j \sim 30-50J$ having in mind tunneling rates in the range $J/2\pi\hbar \sim 20-50$ MHz. We then impose that the constraints required by Floquet and reservoir engineering discussed in the main text are satisfied.   Summarizing, this means
\begin{enumerate}
\item $\hbar\omega \gg J:$ The results shown in the article are obtained choosing $\hbar\omega=20J$. This leads to single-excitation spectra of the Floquet-engineered effective Hamiltonian for the atoms with bandwidth of a few units of $J$.
\item $|\delta_j + U n_j +m\hbar\omega|\gg g_j$. For the given values of $\hbar\omega$ and $U$, the values of the atom-cavity detunings $\delta_j$ are fine tuned by changing the atomic potential $\Delta$ to ensure that Floquet resonances are avoided. 
\item $|\varepsilon_\eta - \varepsilon_{\eta'}| \gg \hbar\kappa_j \gg |\!\bra{\eta}\chi_j(\hat{n}_j)\ket{{\eta'}}\!| \sqrt{\bar{n}_{\mathrm{ph},j}}$: for the engineered dissipative processes to resolve single effective transitions, the effective gaps must be much larger than the decay energy $\hbar\kappa_j$ and than the effective atom-cavity coupling. However, for the cavities to act as a Markovian bath, the decay rate must in turn be much larger than the effective coupling. In few-plaquette systems, the effective gaps are of the order of the effective tunneling rate $J^\mathrm{eff}_{\ell\ell'}\simeq0.5 J$. The effective atom-cavity coupling $\chi_j(\hat{n}_j)$ is a second order process and is in general smaller than these gaps, away from Floquet resonances. Therefore, at fixed $\hbar\kappa_j\ll J$ and given that $d_j$ is fixed by the gap to be controlled, the conditions stated here mainly constrain the pump amplitude $\mathcal{E}_j$ for the cavities. These amplitudes should be as large as possible to enhance the transition rate, but small enough to satisfy the above conditions. In general, we find optimal transition rates for amplitudes leading to $\bar{n}_{\mathrm{ph},j}\sim 1-2$, consistently with what was observed in Ref.~\cite{Hacohen2015} for an undriven system.
\item Another degree of freedom is given by the ``position'' of the cavities, namely to which sites they are coupled. After inspecting the numerical values of the matrix elements of the coupling operators $\bra{\eta} \chi_j(\hat{n}_j)\ket{{\eta'}}$, the choice is done in order to maximize the effective coupling for a desired transition. 
\end{enumerate}
Using these conditions as guiding lines, the search for good parameters is done numerically. An example of this search and of how the dissipative control protocols are designed is illustrated with a small system in the following.

\begin{figure}
    \centering
    \includegraphics[width=\linewidth]{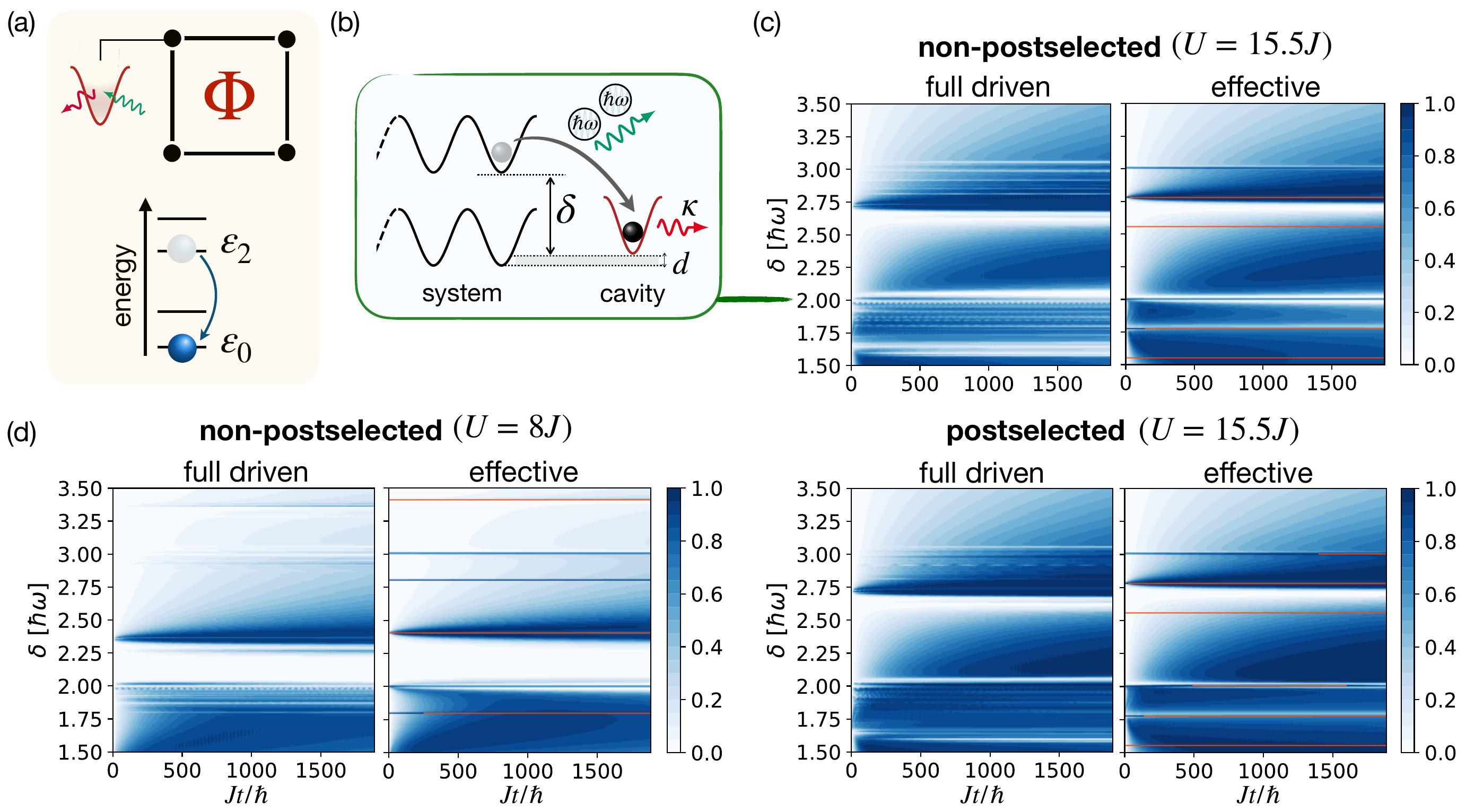}
    \caption{\textbf{(a)} Sketch of the single-plaquette system and dissipative transition induced by the cavity. \textbf{(b)} a process in which one photon tunnels to the cavity by emitting two drive quanta $\hbar\omega$ in the Floquet drive and leaks to the environment, leading to excitation loss. Such processes spoil the efficiency of the Floquet and reservoir engineering as visible from panel (c). \textbf{(c)} effective ground state occupation (color scale) as a function of time and of the atom-cavity detuning $\delta$ for $U=15.5J$. Left and right panels compare the results with the full master equation and the results predicted by the effective master equation, respectively. The other parameters are $\Phi=\pi/2$, $g=J$, $\kappa=0.1J$, $\mathcal{E}=1.6J$, $\hbar\omega=20J$, $\lambda\simeq 0.72 \omega$ (the value of $\lambda$ is such that the effective tunneling strength along every edge is the same, see Section ``Artificial magnetic flux"). The upper panel corresponds to non-postselected results, while the lower panel shows postselected results.  \textbf{(d)} same as in panel (c) for a different interaction value $U=8J$. Thin red lines in the panels ``effective'' in (c) and (d) approximately indicate points where the perturbative expansion diverges, namely they represent regions where one of the denominators in the effective coupling becomes smaller than a numerical threshold $10^{-7}J$.}
    \label{fig:ex1}
\end{figure}

\subsection{Single dissipative transition}

Consider one excitation circulating in an array of four artificial atoms making up a single square plaquette pierced by an artificial magnetic flux $\Phi$ and coupled to a single cavity, as depicted in Fig.~\ref{fig:ex1}(a). In this case, due to the symmetry of the system, the position of the cavity is irrelevant and leads to the same (absolute value of the) matrix elements for the system-cavity coupling. Figure~\ref{fig:ex1} characterizes an ideal transition from the effective second-excited eigenstate $\ket{2}$ of the effective atomic Hamiltonian to the ground state $\ket{0}$. This is realized by setting the cavity-pump detuning to match the energy difference $d = \varepsilon_{2} - \varepsilon_{0}$, then initializing the system in $\ket{2}$ in the simulation and studying the stroboscopic occupation of $\ket{0}$ as a function of time for varying atom-cavity detuning $\delta$, with other parameters fixed.
Panels (c) and (d) compare the prediction of the effective master equation with the full dynamics for two values of the interaction parameter $U$, in (c) also comparing postselected and non-postselected results. First of all, one can see that the effective model, which is much less computationally demanding, can be reliably used for parameter exploration. 
Low-probability (white) lines indicate divergences in the perturbative expansion. These divergences correspond to drive-assisted tunneling between the system and the cavity. For instance, as sketched in panel (b) of Fig.~\ref{fig:ex1}, the white region at $\delta\sim 2\hbar\omega$ in panel (c) corresponds to a process in which the excitation tunnels towards the cavity while emitting two Floquet quanta in the drive. The excitation is eventually lost due to cavity decay.

The effective model is derived using unperturbed gaps that do not include the presence of the tunneling $J$ in the system, which is included instead in the ``diagonal'' perturbation $\hat{\mathcal{V}}_\mathrm{D}$, see Eq.~\eqref{eq:olVD}. The resonances that are visible in the panels ``effective" of Fig.~\ref{fig:ex1}(c)-(d) then split into a larger number of resonances in the full model, which will still be separated by a gap much smaller than $\delta$, since $J\ll \delta$. Moreover, higher-order processes not captured in the effective model will contribute to this effect and further introduce small energy shifts of the resonances. All these effects yield the more complex pattern of ``white resonances'' visible in the panels ``full driven" of~\ref{fig:ex1}(c)-(d). Although these resonances can make the search difficult, good parameter regimes can still be found by carefully avoiding regions where the approximations underlying the effective model break down. 

\subsection{Autonomous stabilization}

\begin{figure}
    \centering
    \includegraphics[width=\linewidth]{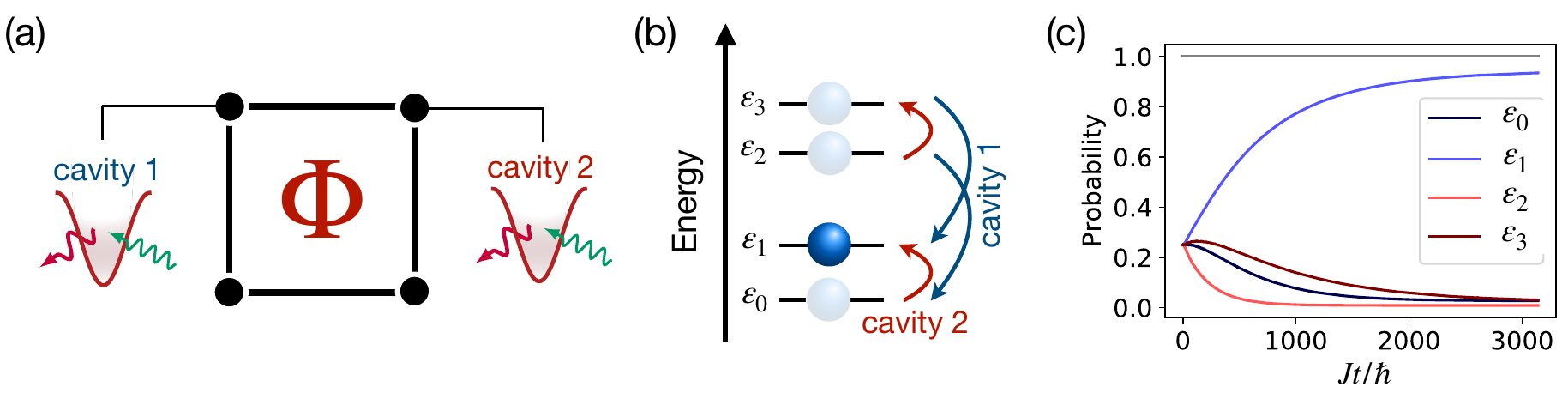}
    \caption{\textbf{Autonomous stabilization.} \textbf{(a)} Two cavities are attached to the upper two sites. \textbf{(b)} Dissipative transitions engineered using the two cavities. The dissipative path pushes the system towards the first excited state of the effective Hamiltonian. \textbf{(c)} Effective-eigenstate populations as a function of time, starting from an infinite-temperature state in the single-excitation manifold. The parameters used are $g_j=J$, $U=15.5J$, $\Delta=2.25\omega$, $\hbar\omega=20J$, $\kappa_j=0.1J$, $\bm{\mathcal{E}}=(1.8J, J)$.}
    \label{fig:auto_stab}
\end{figure}

We now illustrate in the single-plaquette system how protocols for autonomous stabilization are designed. In particular, we stabilize the first-excited state $\ket{1}$ by coupling the system to a second cavity and exploiting both ``down-in-energy'' and ``up-in-energy'' transitions, as depicted in Fig.~\ref{fig:auto_stab}(a)-(b). The first cavity remains set to cool transitions with gap $\varepsilon_{2} - \varepsilon_{0}$, as in the previous subsection. Due to symmetries in the spectrum, the first cavity thus ``cools'' from $\varepsilon_{2}$ to $\varepsilon_{0}$ and from $\varepsilon_{3}$ to $\varepsilon_{1}$. The second cavity-pump detuning is set instead at the (negative) gap $\varepsilon_{0} - \varepsilon_{1}$, thus inducing ``heating'' transitions from $\varepsilon_{0}$ to $\varepsilon_{1}$ and from $\varepsilon_{2}$ to $\varepsilon_{3}$. The fact that $\ket{1}$ is the unique steady state can intuitively be seen by noting that no transition can occur that leaves such a state, while a sequence of transitions can connect any other state to $\ket{1}$.
The stabilization is shown in Fig.~\ref{fig:auto_stab}(c), which depicts the time evolution of the eigenstate occupations (using the full master equation and postselecting -- see Section ``Numerical simulations and postselection'') starting from an infinite-temperature state in the single-excitation subspace, {\it i.e.}, $\hat{\rho}_0 = \frac{1}{M}\sum_{\ell=1}^M \hat{a}_\ell^\dagger\ket{0}\bra{0}\hat{a}_\ell = \mathbb{1}_\mathrm{1p}/M$. The system successfully evolves towards $\ket{{1}}$. Note that such a transition from a fully mixed state to a (nearly) pure state, which entails an entropy decrease, could not be achieved in the absence of dissipative processes.

\end{document}